\documentclass[12pt]{article}
\usepackage{a4wide,epsfig,amsmath,amssymb,cite,scalefnt,graphicx}
\def\prl{Phys.\ Rev.\ Lett.\ }

\def\be{\begin{equation}}
\def\ee{\end{equation}}
\def\besub{\begin{subequations}}
\def\eesub{\end{subequations}}

\def\qbar{\overline q}

\def\gbar{\bar g}
\def\Gbar{\overline G}
\def\Zbar{\overline Z}

\def\pa{\partial}

\def\frak#1#2{{\textstyle{\frac{#1}{#2}}}}

\def\Ncal{{\cal N}}

\def\Fbar{\overline F}

\def\mbar{\overline m}

\def\qbar{\overline q}

\def\Ocal{{\cal O}}
\def\Otcal{\widetilde\Ocal}

\def\hbar{{\overline h}}
\def\sigmabar{{\overline\sigma}}

\def\mbar{\overline{m}}

\def\Phibar{{\overline \Phi}}
\def\thetabar{{\overline\theta}}
\def\alphadot{\dot\alpha}
\def\betadot{\dot\beta}
\def\Qbar{\overline Q}
\def\psibar{\overline{\psi}}

\def\phibar{\overline{\phi}}
\def\ybar{\overline y}
\def\mbar{\overline m}

\newcommand{\beq}{\begin{equation}}
\newcommand{\eeq}{\end{equation}}
\newcommand{\beqa}{\begin{eqnarray}}
\newcommand{\eeqa}{\end{eqnarray}}

\def\lsl{\not{\hbox{\kern-1.7pt $\ell$}}}
\def\ksl{\not{\hbox{\kern-2.1pt $k$}}}
\def\Psl{\not{\hbox{\kern-2.1pt $P$}}}

\def\be{\begin{equation}}
\def\bea{\begin{eqnarray}}
\def\eea{\end{eqnarray}}
\def\nn{\nonumber\\}

\begin{document}

\begin{titlepage}
\begin{flushright}
LTH799\\ 
\end{flushright}

\vspace*{3mm}

\begin{center}
{\Huge
The non-anticommutative supersymmetric Wess-Zumino model}\\[12mm]
{\bf I.~Jack, D.R.T.~Jones\footnote{address Oct 1st-Dec 31st 2008:
TH Division, CERN, 1211 Geneva 23, Switzerland} and R. Purdy}\\

\vspace{5mm}
Dept. of Mathematical Sciences,
University of Liverpool, Liverpool L69 3BX, UK\\

\end{center}

\vspace{3mm}
\begin{abstract}
We discuss the non-anticommutative ($\Ncal=\frak12$) 
supersymmetric Wess-Zumino model in four
dimensions. Firstly we introduce differential operators which implement
the non-anticommutative supersymmetry algebra acting on the component fields 
and action. Then we perform the renormalisation of the model up to two-loop
order, including the complete set of terms necessary for renormalisability.
We show that (at least up to this order)
the results obtained when we eliminate the auxiliary field after
renormalisation are equivalent to those obtained when we eliminate the auxiliary
fields before quantisation.
\end{abstract}

\vfill

\end{titlepage}

\section{Introduction}

The subject of deformed quantum field theories has attracted 
renewed attention in recent years due to their natural appearance 
in string theory. Initial studies were devoted to theories on 
non-commutative spacetime in which the commutator of the spacetime 
co-ordinates becomes
non-zero. More recently\cite{casal}-\cite{oog}, 
non-anticommutative supersymmetric theories have been 
constructed by deforming the anticommutators of the grassmann co-ordinates
$\theta^{\alpha}$
(while leaving the anticommutators of the $\thetabar{}^{\alphadot}$ unaltered).
Consequently, the anticommutators of the supersymmetry generators 
$\Qbar_{\alphadot}$ are
deformed while those of the $Q_{\alpha}$ are unchanged. 
Non-anticommutative versions of the Wess-Zumino model and supersymmetric gauge
theories have been formulated in four 
dimensions\cite{seiberg,araki} and their renormalisability
discussed\cite{brittoa}-\cite{lunin}, with explicit computations up to two 
loops\cite{grisa} for the Wess-Zumino 
model and one loop for gauge theories\cite{jjwa}--\cite{jjwc}. 

More recently 
still, 
non-anticommutative theories in two dimensions have been considered. On the 
one hand, non-anticommutative versions of particular non-linear $\sigma$-models 
have been constructed (by dimensional reduction from four 
dimensions)\cite{inami} and the one-loop corrections computed\cite{arakib}; 
on the 
other hand, a non-anticommutative version of the general $\Ncal=2$ 
K\"ahler $\sigma$-model has been constructed directly in two dimensions, 
initially in Refs.~\cite{chand, chanda} but then given an elegant
reformulation in Refs.~\cite{luis,chandc}. The one-loop divergences for this
model were computed in Ref.~\cite{jp}, where it was found convenient to
introduce differential operators implementing the deformed supersymmetry 
algebra acting on the component fields. 

In this article we return to a closer examination of the non-anticommutative
Wess-Zumino model in four dimensions. Firstly, we show how analogues of the 
differential operators introduced in Ref.~\cite{jp} may easily be constructed 
to implement the non-anticommutative supersymmetry algebra for this model 
in its component formulation. Next we re-examine the two-loop calculation
first performed in Ref.~\cite{grisa}, showing that to correctly obtain  
results for the theory where the auxiliary fields have been eliminated   
from those for the uneliminated theory, it is necessary to include
in the classical action separate couplings for all the terms which may be 
generated by the renormalisation process.  
We show that (at least up to two-loop order)
the results obtained when we eliminate the auxiliary field after
renormalisation are equivalent to those obtained when we eliminate the auxiliary
fields before quantisation.

\section{Representation of the supersymmetry algebra; the undeformed case}
We follow the analysis of Ref.~\cite{jp} in determining differential operators
which represent the supersymmetry algebra in the undeformed and deformed 
cases. The supersymmetry charges are 
\be
Q_{\alpha}=\frac{\pa}{\pa\theta^{\alpha}},\quad \Qbar_{\alphadot}=
-\frac{\pa}{\pa\thetabar{}^{\alphadot}}+2i\theta^{\alpha}
\pa_{\alpha\alphadot},
\ee
where
\bea
\pa_{\alpha\alphadot}&=&\sigma^{\mu}_{\alpha\alphadot}{\pa\over{\pa y^{\mu}}},
\nn
y^{\mu}&=&x^{\mu}+i\theta^{\alpha}\sigma^{\mu}_{\alpha\alphadot}
\thetabar{}^{\alphadot}.
\label{eq:ydef}
\eea
They satisfy the algebra
\bea
\{Q_{\alpha},Q_{\beta}\}=0,&\qquad&
\{\Qbar_{\alphadot},\Qbar_{\betadot}\}
=0,\nn
\{\Qbar_{\alphadot},Q_{\alpha}\}&=&2i\pa_{\alpha\alphadot}.\label{eq:undefcom}
\eea
The superfields have expansions in terms of component fields given by
\bea
\Phi&=&\phi+\sqrt2\theta\psi+\theta^2 F,\\
\Phibar&=&\phibar
+\sqrt2\thetabar\psibar-2i\theta\sigma^{\mu}\thetabar\pa_{\mu}\phibar
+\thetabar^2\left[\Fbar
+i\sqrt2\theta\pa\psibar 
+\theta^2\pa^2\phibar\right],
\label{compexp}
\eea 
where the component fields are functions of $y^{\mu}$, as defined in
Eq.~(\ref{eq:ydef}).
It is useful to represent the charges $Q_{\alpha}$, $\Qbar_{\alphadot}$
by differential operators $q_{\alpha}$, $\qbar_{\alphadot}^0$ acting
on the fields, i.e.
\bea
\left[Q_{\alpha},\Phi\right]&=&q_{\alpha}\Phi,\nn
\left[\Qbar_{\alphadot},\Phi\right]&=&\qbar{}^0_{\alphadot}\Phi
\label{eq:undefopa}
\eea
(with similar expressions for $\Phibar$) where
\besub
\bea
{1\over{\sqrt2}}q_{\alpha}&=&
\psi_{\alpha}\frac{\pa}{\pa\phi}+F\frac{\pa}{\pa\psi^{\alpha}}
-i[\pa_{\alpha\alphadot}\phibar]\frac{\pa}{\pa\psibar_{\alphadot}}+i
[\pa_{\alpha\alphadot}\psibar^{\alphadot}]
\frac{\pa}{\pa\Fbar},\label{eq:undefopba}\\
{1\over{\sqrt2}}\qbar{}_{\alphadot}^0&=&
\psibar_{\alphadot}\frac{\pa}{\pa\phibar}-\Fbar\frac{\pa}
{\pa\psibar^{\alphadot}}
-i[\pa_{\alpha\alphadot}\phi]\frac{\pa}{\pa\psi_{\alpha}}
-i[\pa_{\alpha\alphadot}\psi^{\alpha}]\frac{\pa}{\pa F}.
\label{eq:undefopbb}
\eea
\eesub
The superscript ``0'' is in anticipation of a different form for 
$\qbar_{\alphadot}$ in the deformed case; while $q_{\alpha}$, on the other hand,
will be unchanged.
Our convention will be that a derivative or other operator acts on 
everything to its right, unless enclosed in square brackets.
(By the way, note that
\be
\epsilon_{\alpha\beta}{\pa\over{\pa\psi_{\beta}}}
=-{\pa\over{\pa\psi^{\alpha}}}.)
\ee
 
\section{Representation of the supersymmetry algebra; the deformed case}
In this section we repeat the analysis of the previous section for the case
of deformed supersymmetry.
For the deformed version we take
\be
\{\thetabar{}^{\alphadot},\thetabar{}^{\betadot}\}=0,
\quad \{\theta^{\alpha},\theta^{\beta}\}=C^{\alpha\beta}.
\ee
The charges then satisfy the algebra
\bea
\{Q_{\alpha},Q_{\beta}\}=0,&\qquad&
\{\Qbar_{\alphadot},\Qbar_{\betadot}\}
=-4C^{\alpha\beta}\pa_{\alpha\alphadot}
\pa_{\beta\betadot},\nn
\{\Qbar_{\alphadot},Q_{\alpha}\}&=&2i\pa_{\alpha\alphadot}.\label{eq:acomm}
\eea
The non-anticommutativity is implemented at the level of
superfields by introducing the Moyal $*$-product, which satisfies
\bea
\theta^{\alpha}*\theta^{\beta}&=&-\frac12\epsilon^{\alpha\beta}\theta^2
+\frac12C^{\alpha\beta},\nn
\theta^{\alpha}*\theta^2&=&C^{\alpha\beta}\theta_{\beta},\nn
\theta^2*\theta^2&=&-\det C\equiv{1\over {M^2}}.
\eea
We now wish to construct differential operators $\qbar_{\alphadot}$ representing
the effects of $\Qbar_{\alphadot}$ in the deformed case in a similar manner to
Eq.~(\ref{eq:undefopa}),
extending $\qbar_{\alphadot}^0$ given in Eq.~(\ref{eq:undefopbb}) for the 
undeformed case. (As mentioned before, 
the operators $q_{\alpha}$ are unchanged by the deformation.)
We start by examining the effects of $\Qbar_{\alphadot}$ on powers of
$\Phi$ alone, since mixed products of $\Phi$ and $\Phibar$ present
additional complications. Defining
\be
I_r^{(n)}(\phi,F)=\int_{-\frac12}^{\frac12}d\xi\left(\frac{\xi}{2M}\right)^r
\left(\phi+\frac{2\xi}{M}F\right)^n
\ee
it is straightforward to show using the methods of Ref.~\cite{luis} that
\be 
\Phi^n_*=
(1+\theta q -\frak14\theta^2 q^2)\left(I_0^{(n)}-q^2I_1^{(n)}\right),
\label{suppot}
\ee
where $\Phi^n_*$ denotes the $*$-product of $n$ $\Phi$'s.
Then acting on $\Phi^n_*$, $\Qbar_{\alphadot}$ is represented by
\be
\qbar^{\Phi}_{\alphadot}=\qbar{}^{0}_{\alphadot}
-i(qC\pa)_{\alphadot}+4i\left(
-q^{2}[\pa_{\alpha\alphadot}q^{\alpha}]\Otcal
+\pa_{\alpha\alphadot}q^{\alpha}\Ocal+
[\pa_{\alpha\alphadot}q^{\alpha}]\Ocal\right). \label{eq:fullq}
\ee
Here
\bea
\Ocal I_0^{(n)}&=&I_1^{(n)},\nn
\Ocal I_1^{(n)}&=& I_2^{(n)}-\Otcal I_0^{(n)}.\label{eq:orels}
\eea
These properties are guaranteed by the following definitions:
\bea
\Ocal&=&\sum_{r=1}^{\infty}a_r
\left(\frac{1}{4M^2}\right)^r
\left(4F\frac{\pa}{\pa \phi}\right)^{2r-1},\nn
\Otcal&=&\sum_{r=1}^{\infty}(2r-1)a_r
\left(\frac{1}{4M^2}\right)^r
\left(4F\frac{\pa}{\pa \phi}\right)^{2r-2},\eea
where the $a_r$ must satisfy for each $n\ge1$\cite{jp}
\be
\sum_{r=0}^{n-1}\frac{a_{n-r}}{2^{2r}(2r+1)(2r)!}=\frac{1}{
2^{2n}(2n+1)(2n-1)!},
\ee 
the first few being given by 
\be
a_1=\frac{1}{12}, \quad a_2=-\frac{1}{720},\quad a_3=\frac{1}{2^5.3^3.5.7}.
\ee
To check that the operators in Eq.~(\ref{eq:fullq}) do indeed
represent the operators $\Qbar_{\alphadot}$ according to
\be
[\Qbar_{\alphadot},\Phi^n_*]_*=\qbar^{\Phi}_{\alphadot}\Phi^n_*
\ee
(where $[\phantom{q},\phantom{q}]_*$ represents the commutator evaluated using
$*$-products)
we need to use Eqs.~(\ref{eq:orels}) in
conjunction with
\be
\qbar_{\alphadot}^{0}I_r^{(n)}
=-i[\pa_{\alpha\alphadot}
q^{\alpha}]I_{r+1}^{(n)}.\label{eq:idsa}
\ee

Of course it is not sufficient to reproduce the effects of
$\Qbar_{\alphadot}$ on $*$-products of $\Phi$ alone; we saw in the 
two-dimensional case that it was necessary to consider the 
effect on deformed versions of general polynomials in $\Phi$ and $\Phibar$,
such as the K\"ahler potential. In the case of the four-dimensional 
Wess-Zumino model, to investigate 
the divergence structure it would be sufficient to consider only the effects
of $\Qbar_{\alphadot}$ on cubic superpotentials in either $\Phi$ or $\Phibar$,
or on $\Phi*\Phibar$. However, in general, if one were interested in 
contributions to the effective action, one would need once again to consider 
deformed versions of general polynomials; and so we shall again take the 
K\"ahler potential as an example.

For an undeformed K\"ahler potential
\be
K[\Phi,\Phibar]=\sum_{n,m}K_{n,m}\Phi^n\Phibar{}^m,
\ee 
the natural definition of the deformed K\"ahler potential is
\be
K_*[\Phi,\Phibar]=\sum_{n,m}K_{n,m}[\Phi^n\Phibar{}^m]_*,
\ee
where $[\Phi^n\Phibar{}^m]_*$ represents the symmetrised $*$-product of
$n$ $\Phi$'s and $m$ $\Phibar$'s. It can be shown that
\bea
K_*[\Phi,\Phibar]&=&
(1+\theta q-\frak14\theta^2 q^2)\left[1+\thetabar
\qbar^{0\prime\prime}
-\frak14\thetabar^2 (\qbar^{0\prime\prime})^2\right]\nn
&&\left[K_0(\phi,F,\phibar)-q^2K_1(\phi,F,\phibar)\right]\nn
&&-\frac{1}{4M^2}\thetabar^2 q^{\prime2}(\pa'')^2
K_0(\phi,F,\phibar),
\label{Kstar}
\eea
where
\be 
K_m(\phi,F,\phibar)=\int_{-\frac12}^{\frac12}d\xi\left(\frac{\xi}{2M}\right)^m
K\left(\phi+
\frac{2\xi}{M}F,\phibar\right).
\ee
Here a prime denotes the part of the operator containing derivatives
with respect to the chiral (but not the anti-chiral) fields, and
correspondingly 
a double prime denotes the part of the operator containing derivatives
with respect to the anti-chiral (but not the chiral) fields.
Moreover,
\be 
\pa_{\mu}''=[\pa_{\mu}\phibar]\frac{\pa}{\pa\phibar}+
[\pa_{\mu}\psibar_{\alphadot}]\frac{\pa}{\pa\psibar_{\alphadot}}
+[\pa_{\mu}\Fbar]\frac{\pa}{\pa \Fbar},
\ee
with a similar expression for $\pa_{\mu}'$.
The corresponding version of the operator representing $\Qbar_{\alphadot}$ is
$\qbar_{\alphadot}$ defined by 
\bea 
\qbar_{\alphadot}&=&\qbar{}^{0}_{\alphadot}
-i(qC\pa)_{\alphadot}-\frac{i}{4M^2}
\left(\pa''_{\alpha\alphadot}q^{\prime\prime\alpha}q^{\prime2}
+\pa'_{\alpha\alphadot}q^{\prime\alpha}q^{\prime\prime2}\right)\nn
&&+4i\left(
-q^{\prime2}[\pa'_{\alpha\alphadot}q^{\prime\alpha}]\Otcal
+\pa'_{\alpha\alphadot}q^{\prime\alpha}\Ocal+
[\pa'_{\alpha\alphadot}q^{\prime\alpha}]\Ocal\right).
\label{eq:fullqa}   
\eea
We can verify that these
operators do indeed implement the operators $\Qbar_{\alphadot}$ according to
\be
[\Qbar_{\alphadot},K_*]_*=\qbar_{\alphadot}K_*,
\ee
using
the analogue of Eq.~(\ref{eq:orels}) for the K\"ahler potential,
\bea
\Ocal K_0&=&K_1,\nn
\Ocal K_1&=& K_2-\Otcal K_0\label{eq:orelsa}
\eea
together with the analogue of Eq.~(\ref{eq:idsa}),
\be
\qbar_{\alphadot}^{\prime 0}K_r=-i[\pa'_{\alpha\alphadot}q^{\prime\alpha}]
K_{r+1},\quad
q^{\prime\prime}_{\alpha}K_0=q^{\prime\prime}_{\alpha}K_1=0.
\ee 

Moreover, it is easy to check that the operators $q_{\alpha}$
in Eq.~(\ref{eq:undefopba}) and $\qbar_{\alphadot}$ in Eq.~(\ref{eq:fullqa})
satisfy the anticommutation relations of Eq.~(\ref{eq:acomm}),
using
\be 
\left[\qbar^{0}_{\alphadot},F\frac{\pa}{\pa \phi}\right]
=-i[\pa'_{\alpha\alphadot}q'^{\alpha}]
\ee
(which implies
\be 
[\qbar_{\alphadot}^0,\Ocal]
=-i[\pa'_{\alpha\alphadot}q'^{\alpha}]\Otcal).
\ee

The kinetic part of the standard Wess-Zumino model may be obtained as in 
Ref.~\cite{jp} from 
Eq.~(\ref{Kstar}) taking simply $K[\Phi,\Phibar]=\Phi\Phibar$, 
\bea
S_{\rm kin}&=&\int d^4xd^2\theta d^2\thetabar\ \Phi*\Phibar
=\frak{1}{16}\int d^4x\ q^2\qbar^2\phi\phibar\nn
&=&\int d^4x
\Bigl(\pa^{\mu}\phibar\pa_{\mu}\phi+i\psibar\sigmabar^{\mu}\pa_{\mu}\psi 
+\Fbar F\Bigr),
\label{inva}
\eea
so we see that the kinetic terms are undeformed.
We see from Eq.~(\ref{suppot}) that the holomorphic potential terms are 
given by
\be
S_{W}
=-\int d^4x\int d^2\theta\ [\frak12m \Phi_*^2+\frak16y\Phi_*^3]
=\frak14\int d^4x\ q^2[\frak12mI_0^{(2)}+\frak16yI_0^{(3)}]
\label{invb}
\ee
which leads to 
\be
S_{W}=\int d^4x\ [\frak12m(\psi^2-F\phi)+\frak12y(\phi\psi^2-F\phi^2)+
\frak16y(\det C) F^3].
\ee
Since $\Phibar^n_*=\Phibar^n$, 
the antiholomorphic potential terms are given by
\be
S_{\overline W}
=-\int d^4x\int d^2\thetabar\ [\frak12\mbar \Phibar^2+\frak16\ybar\Phibar^3]
=\frak14\int d^4x\ \qbar^2[\frak12\mbar\phibar^2+\frak16\ybar\phibar^3]
\label{invc}
\ee
which leads to
\be
S_{\overline W}=
\int d^4x\ [\frak12\mbar(\psibar^2-\Fbar\phibar)+\frak12\ybar(\phibar\psibar^2
-\Fbar\phibar^2)],
\ee
with no deformation (the deformed part of $\qbar$ in 
Eq.~(\ref{eq:fullqa}) having no effect on a function of $\phibar$).  
The full classical $\Ncal=\frac12$ action is therefore
\bea
S&=&\int d^4x 
\Bigl(\pa^{\mu}\phibar\pa_{\mu}\phi+i\psibar\sigmabar^{\mu}\pa_{\mu}\psi 
+\Fbar F  -GF-\Gbar\Fbar\nn
&&+\frak12y\phi\psi^2+\frak12\ybar\phibar\psibar^2+\frak12m\psi^2
+\frak12\mbar\psibar^2
+\frak16 y(\det C) F^3\Bigr),
\label{defact}
\eea
where $G=m\phi+\frak12y\phi^2$. 
This action was first derived in Ref.~\cite{seiberg} by taking the standard
undeformed ($\Ncal=1$) action in superfields and replacing ordinary products 
by Moyal $*$-products. 

In the undeformed case, expressions like those in Eqs.~(\ref{inva}),
(\ref{invb}) and (\ref{invc})
in terms of $q_{\alpha}$ and $\qbar^0_{\alphadot}$ 
encapsulate the supersymmetry of the undeformed action $S_0$ 
due to the nilpotency 
of $q_{\alpha}$, $\qbar_{\alphadot}$ and the fact that $q_{\alpha}$, 
$\qbar_{\alphadot}^0$
annihilate functions of $\phibar$, $\phi$, respectively, leading to
\be
q_{\alpha}S_0=\qbar^0_{\alphadot}S_0=0.
\ee 
In the deformed case, although 
\be
q_{\alpha}S_{\rm kin}=\qbar_{\alphadot} S_{\rm kin} = q_{\alpha}S_{W}=
q_{\alpha}S_{\overline W} =\qbar_{\alphadot} S_{\overline W}=0,
\ee
it is not the case that $\qbar_{\alphadot} S_{W}=0$; 
indeed it is no longer the case that $\qbar^0_{\alphadot} S_{W}=0$ either. 
It is only the tranformations generated by $Q_{\alpha}$ 
that are a symmetry of the deformed action with potential. It is worth 
mentioning that in the two-dimensional case (where we considered only the 
kinetic part of the action derived from the K\"ahler potential), despite our 
classical action being annihilated by $q$ and $\qbar$, the divergent quantum
corrections were only annihilated by $q$. In fact it is only for simple
linear field transformations that one can prove that an invariance of the 
classical action results in a similar invariance of the quantum corrections;
for non-linear gauge transformations the quantum invariance is encapsulated in 
Ward identities, and for transformations such as our deformed 
$\qbar_{\alphadot}$ some
other formulation may be possible.

\section{Renormalisation}
In this section we discuss the renormalisation of the 
non-anticommutative Wess-Zumino model up to two loops. Two-loop calculations
for this model were first performed by Grisaru et al\cite{grisa}; here we 
extend their calculation by including from the outset the full set of terms 
which can be generated by renormalisation. We shall postpone until later a
detailed comparison of our results with theirs. 

The only effect of the non-anticommutativity in the component action 
of Eq.~(\ref{defact}) is the
final $(\det C) $ term. As we emphasised at the end of the last section,
the deformed action is only invariant under the transformations
generated by $Q_{\alpha}$. 
This term is in fact separately invariant under these 
residual $\Ncal=\frac12$ transformations 
and so there is no reason for the
coefficient to evolve in the same manner as the Yukawa coupling in the 
$\Ncal=1$ part of the action; and therefore we are at liberty (possibly even 
obliged) 
to introduce this term with its own separate coefficient. In fact, 
this term generates one-loop
divergences whose cancellation requires
$(\det C) F^2\Gbar$ and $\mbar^2(\det C) F^2$ 
terms in the action; and these terms in turn generate further $(\det C) F^3$
divergences together with other new terms. All these terms should be included
in the classical action with their own coefficients in order to guarantee
renormalisability. 
It is easy to see 
which additional terms can be generated\cite{lunin}. (For a complete 
analysis in the general, gauged case, see Ref.~\cite{jjwc}.) The
action has a ``pseudo R-symmetry'' under 
\be
\phi\rightarrow e^{-i\omega}\phi,\quad
F\rightarrow e^{i\omega}F,\quad
C^{\alpha\beta}\rightarrow e^{-2i\omega}C^{\alpha\beta},\quad
y\rightarrow e^{i\omega}y,
\label{psea}
\ee
$\Fbar$, $\phibar$ and $\ybar$
transforming with opposite charges to $F$, $\phi$ and $y$ respectively,
and $\psi$, $\psibar$ being
neutral; and also a ``pseudo-chiral symmetry'' under 
\be
\phi\rightarrow e^{i\gamma}\phi,\quad
m\rightarrow e^{-2i\gamma}m,\quad
y\rightarrow e^{-3i\gamma}y,
\label{pseb}
\ee
$F$ and $\psi$ transforming in a similar fashion to $\phi$ and barred 
quantities transforming with opposite charges.
 The divergent terms which can arise subject to these invariances have
been enumerated\cite{lunin} and consist (for the ungauged case) of 
\bea
\ybar^{-1}(\det C) F^3, &\quad&\ybar^{-1}(\det C) F^2\Gbar{}, \quad \quad
\ybar^{-1}(\det C) F\Gbar{}^2, \nn
\ybar^{-1}(\det C) \Gbar{}^3, &\quad&\ybar^{-2}\mbar^2(\det C) F^2, 
\quad \quad\ybar^{-2}\mbar^2(\det C) F\Gbar, \nn
\ybar^{-2}\mbar^2(\det C) \Gbar{}^2, &\quad& \ybar^{-3}\mbar^4 (\det C) F,
\quad \quad \ybar^{-3}\mbar^4(\det C) \Gbar.
\label{terms}
\eea
We have anticipated here the fact, adumbrated in Ref.~\cite{grisa}
and proved in Ref.~\cite{brittob}, that the divergences form combinations
of $F$ and $\Gbar$; so that we need only a single coupling 
to remove divergences in (for instance) $(\det C) F^2\ybar\phibar^2$ and
$(\det C) F^2\mbar\phibar$.
We have included in (\ref{terms})
the appropriate factors of $\ybar$ for invariance
under the pseudo-chiral symmetry. These factors are not uniquely determined 
since $y\ybar$ is invariant under this symmetry; the choice we have made
is both concise and motivated by later considerations.  
Each of these terms is separately 
$\Ncal=\frac12$ invariant and so there is nothing in the classical theory to 
determine their coefficients; but we shall investigate whether 
renormalisability has anything to tell us about their values. In fact, some of
the terms listed in (\ref{terms}) could be omitted and still leave a
renormalisable theory (at least up to two-loop order); but nevertheless for
completeness we 
shall include
all the above terms in our classical action. (As pointed out in 
Ref.~\cite{jjwc}, terms of the form $yF\psi C\psi$ and $\phibar^2\psi C\psi$
are possible; but not in the present ungauged case with only one chiral 
field.)

We are therefore led to the action 
\bea
S&=&\int d^4x 
\Bigl( \pa^{\mu}\phibar 
\pa_{\mu}\phi+i\psibar\sigmabar^{\mu}\pa_{\mu}\psi+\Fbar F-GF-\Gbar\Fbar\nn
&&+\frak12y\phi\psi^2+\frak12\ybar\phibar\psibar^2+\frak12m\psi^2
+\frak12\mbar\psibar^2\nn
&&+\ybar^{-1}\left[\frak16 k_1F^3+\frak12k_2F^2\Gbar+\frak12k_3F\Gbar{}^2+
\frak16k_4\Gbar{}^3\right]\nn
&&+\frak12\ybar^{-2}\mbar^2\left[k_5F^2+2k_6F\Gbar+k_7\Gbar{}^2\right]+
\ybar^{-3}\mbar^4[k_8F+k_9\Gbar]
\Bigr).
\label{sunel}
\eea
 Here the $(\det C) $ has been absorbed into
the coefficients $k_{1-9}$. We note that we have no way to determine 
the renormalisation of $(\det C) $ separately, only that of the coefficients
$k_{1-9}$. 

\begin{figure}[H]
\includegraphics{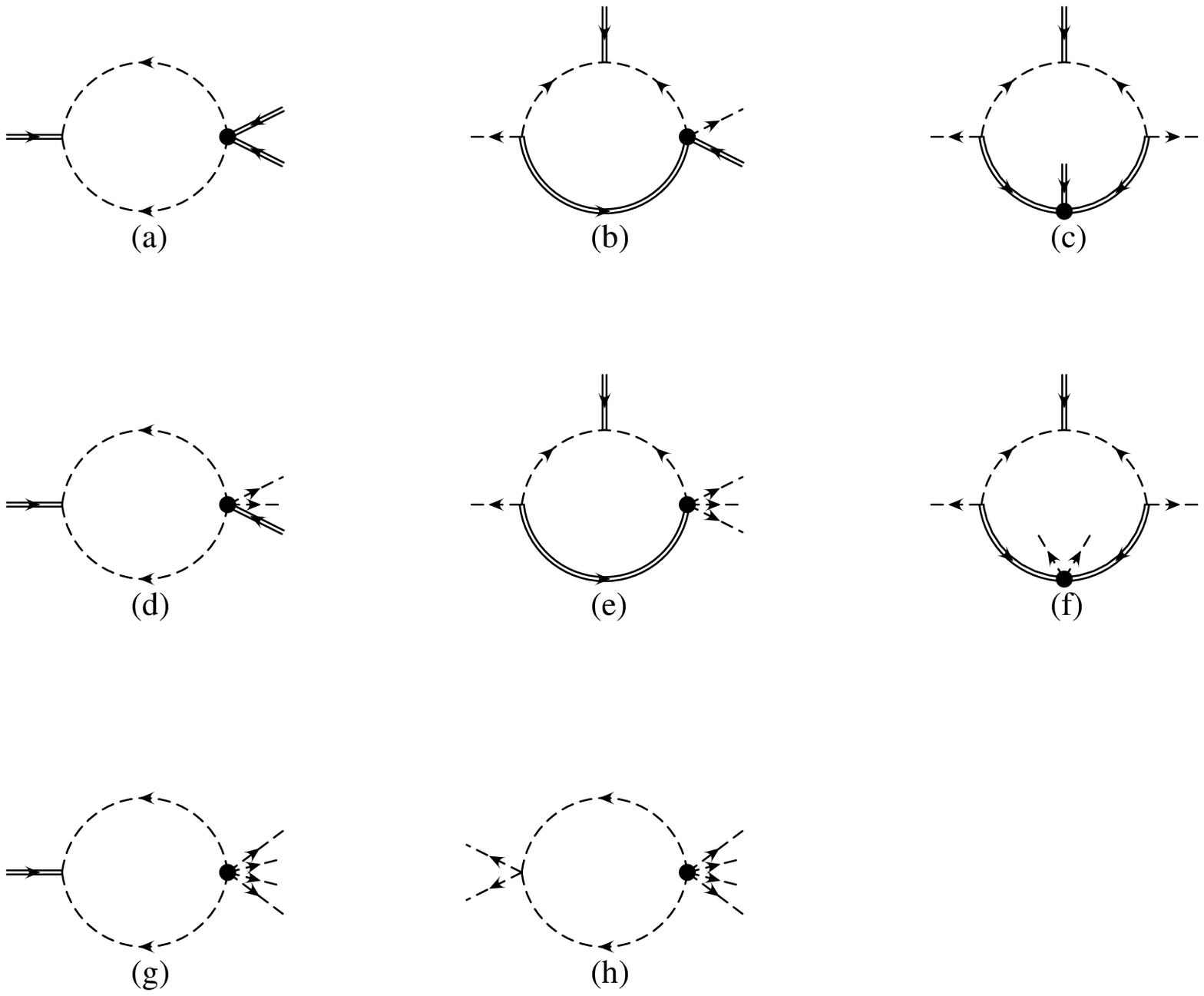}
\caption{One-loop Diagrams (dashed, full, double lines representing
scalar, fermion and auxiliary fields respectively)}\label{fig1}
\end{figure}

We write the divergent contributions to the deformed part of the 
effective action in the form
\bea
\Gamma^{\rm pole}_C&=&-\int d^4x 
[\ybar^{-1}(\Zbar_1F^3+\Zbar_2F^2\Gbar+\Zbar_3F\Gbar{}^2+\Zbar_4\Gbar{}^3)\nn
&&+\ybar^{-2}\mbar^2\left(\Zbar_5F^2+\Zbar_6F\Gbar+\Zbar_7\Gbar{}^2\right)
+\ybar^{-3}\mbar^4(\Zbar_8F+\Zbar_9\Gbar)].
\label{divdef}
\eea
(Note the overall minus sign, introduced to avoid a proliferation of negative 
signs later on.) Note also
that $\Zbar_{1-9}$ will be assumed to contain no finite parts.  
The divergent diagrams contributing to $\Zbar_{1-9}$ can be divided
into groups, each group containing diagrams which have the same 
internal lines and numbers of vertices and also the same number
of external $F$ lines; thus only differing in the numbers of
external $\phibar$ lines and attendant $\ybar$ or $\mbar$ couplings at the 
vertices. The divergent contributions within each group can be expressed 
purely in terms of $F$ and $\Gbar$.
In Fig.~\ref{fig1} are depicted examples of each group at one loop (the one
with the maximal number of external $\phibar$ lines). 
Their divergent contributions are shown diagram by
diagram in Table \ref{taba} and given in total by
\bea
\Zbar_1^{(1)}&=&\frak12k_2\frac{L}{\epsilon},\nn
\Zbar_2^{(1)}&=&(2k_1+4k_2+3k_3)\frac{L}{\epsilon},\nn
\Zbar_3^{(1)}&=&\left(2k_2+4k_3+\frak52k_4\right)\frac{L}{\epsilon},\nn
\Zbar_4^{(1)}&=&0,\nn
\Zbar_5^{(1)}&=&\left(k_1+2k_2+k_3
+k_6\right)\frac{L}{\epsilon},\nn
\Zbar_6^{(1)}&=&\left(k_2+2k_3+k_4
+2k_5+4k_6+3k_7\right)\frac{L}{\epsilon},\nn
\Zbar_7^{(1)}&=&0,\nn
\Zbar_8^{(1)}&=&(k_5+2k_6+k_7+k_9)\frac{L}{\epsilon},\nn
\Zbar_9^{(1)}&=&0,
\label{zonea}
\eea
where
\be
L=\frac{y\ybar}{16\pi^2}.
\ee
\begin{table}
\begin{center}
\begin{tabular}{|c| c c c c c c|} \hline
&$\Zbar_1$& $\Zbar_2$& $\Zbar_3$&$\Zbar_5$&$\Zbar_6$&$\Zbar_8$  \\ \hline
a&$\frak12k_2$ & &&  
&& \\ \hline
b&&$4k_2$& & $2k_2$&
  &  \\ \hline
c&&$2k_1$&&$k_1$&&\\ \hline
d&&$3k_3$&&$k_3+k_6$&&\\ \hline
e&&&$4k_3$&&$2(k_3+2k_6)$&$2k_6$\\ \hline
f&&&$2k_2$&&$k_2+2k_5$&$k_5$\\ \hline
g&&&$\frak52k_4$&&$k_4+3k_7$&$k_7+k_9$ \\ \hline
\end{tabular}
\caption{\label{taba} Divergent contributions from Fig.~\ref{fig1}}
\end{center}
\end{table}
(In Table \ref{taba} the factors of $\frak{L}{\epsilon}$ are suppressed.)
These divergences are cancelled as usual by replacing the parameters
$y$, $\ybar$, $k_{1-9}$ and the fields $\phi$, $\phibar$, $F$, $\Fbar$,
$\psi$, $\psibar$ by corresponding appropriately-chosen bare quantities 
$y_B$, $\ybar_B$, $k_{1B-9B}$, $\phi_B$, $\phibar_B$, $F_B$, $\Fbar_B$,
$\psi_B$, $\psibar_B$, with the bare fields $\phi_B$, $\phibar_B$,
$\psi_B$, $\psibar_B$, given by
$\phi_B=Z^{\frak12}\phi$, etc. (In the case of 
the simple Wess-Zumino model, the same $Z$ is used for each bare field.)
However, there is a subtlety relating to $F_B$, $\Fbar_B$. In the case of 
the gauged $\Ncal=1$ and $\Ncal=\frak12$ Wess-Zumino models we found it 
necessary to make non-linear renormalisations of $F$, $\Fbar$ in order to
ensure multiplicative renormalisability. Here it is not
obligatory but nonetheless we shall explore the freedom of making such 
renormalisations, which will introduce an arbitrariness in the 
$\beta$-functions for $k_i$. Specifically we shall put
\bea
F_B&=&Z^{\frak12}F,\nn
\Fbar_B&=&Z^{\frak12}\Fbar+
\ybar_B^{-1}(\frak12R_BF_B^2+S_BF_B\Gbar_B+\frak12T_B\Gbar_B^2)\nn
&&+\mbar_B^2(U_BF_B+V_B\Gbar_B)+\ybar_B\mbar_B^4 W_B,
\label{fren}
\eea     
where $R_B$, $S_B$, $T_B$, $U_B$, $V_B$ and $W_B$ contain divergent 
contributions only.
 
Moreover, the non-renormalisation theorem leads to 
\be
y_B=\mu^{\frac12\epsilon}Z^{-\frac32}y, \quad\ybar_B=
\mu^{\frac12\epsilon}Z^{-\frac32}\ybar,
\quad m_B=Z^{-1}m,\quad \mbar_B=Z^{-1}\mbar,
\label{nonren}
\ee 
where $\mu$ is the usual dimensional regularisation mass parameter, and hence
\be
\Gbar_B=Z^{-\frac12}\left(\mu^{\frac12\epsilon}\ybar\phibar^2
+\mbar\phibar\right).
\label{Gren}
\ee
Writing  
\bea
k_{iB}&=&k_i+\sum_{n=1} k_{iB}^{(n)},\quad i=1\ldots4,\nn
k_{iB}&=&\mu^{-\epsilon}\left(k_i+\sum_{n=1} k_{iB}^{(n)}\right),
\quad i=5,6,7,\nn
k_{iB}&=&\mu^{-2\epsilon}\left(k_i+\sum_n k_{iB}^{(n)}\right),
\quad i=8,9,
\label{eq1}
\eea
noting that $(\det C) $ has dimension 2, and where
$n$ counts the loop order, and 
\be
k_{iB}^{(n)}=
\sum_{m=1}^n\frac{\kappa^{(n,m)}_i}{\epsilon^m},
\quad i=1\ldots9,
\label{eq1a}
\ee
we find (from the bare version of Eq.~(\ref{sunel}))
\bea
k_{1B}^{(1)}&=&6\Zbar_1^{(1)}-3Z^{(1)}k_1-3R^{(1)}_B,\nn
k_{2B}^{(1)}&=&2\Zbar_2^{(1)}-2Z^{(1)}k_2+(R^{(1)}_B-2S^{(1)}_B),\nn
k_{3B}^{(1)}&=&2\Zbar_3^{(1)}-Z^{(1)}k_3+(2S^{(1)}_B-T^{(1)}_B),\nn
k_{4B}^{(1)}&=&6\Zbar_4^{(1)}+3T^{(1)}_B,\nn
k_{5B}^{(1)}&=&2\Zbar_5^{(1)}-2Z^{(1)}k_5-2U^{(1)}_B,\nn
k_{6B}^{(1)}&=&\Zbar_6^{(1)}-Z^{(1)}k_6+(U^{(1)}_B-V^{(1)}_B),\nn
k_{7B}^{(1)}&=&2\Zbar_7^{(1)}+2V^{(1)}_B,\nn
k_{8B}^{(1)}&=&\Zbar_8^{(1)}-Z^{(1)}k_8-W^{(1)}_B,\nn
k_{9B}^{(1)}&=&\Zbar_9^{(1)}+W^{(1)}_B.
\label{kone}
\eea
($R_B$, $S_B$, $T_B$ have similar expansions to $k_{1B-4B}$ in Eqs.~(\ref{eq1}),
(\ref{eq1a}), but with, for example,
\be
R_B=\sum_{m=1}^n\frac{r^{(n,m)}}{\epsilon^m};
\label{eq1b}
\ee
while $U_B$, $V_B$ have similar expressions to $k_{5B-7B}$
and $W_B$ to $k_{8B,9B}$.)  
We then find (writing $\beta_i=\mu\frac{d}{d\mu}k_i$ and as usual requiring
that $k_{iB}$ in Eq.~(\ref{eq1}) be independent of $\mu$) that
\be
\beta^{(1)}_i=\kappa^{(1,1)}_i,\quad i=1\ldots9.
\label{eq5}
\ee
The $\beta$-functions for $y$, $\ybar$ are defined similarly;
we have 
\be
Z^{(1)}=-\frac{L}{\epsilon}
\label{zone}
\ee
and then by virtue of Eq.~(\ref{nonren})
\be
\beta^{(1)}_y=\frac32Ly,
\label{betaone}
\ee
with a similar expression for $\beta^{(1)}_{\ybar}$. 
From Eqs.~(\ref{zonea}), (\ref{eq1}--\ref{zone}), and writing
$r^{(1,1)}=r_1L$, etc, we have
\bea
\beta^{(1)}_1=\kappa^{(1,1)}_1&=&3(k_1+k_2-r_1)L,\nn
\beta^{(1)}_2=\kappa^{(1,1)}_2&=&(4k_1+10k_2+6k_3+r_1-2s_1)L,\nn
\beta^{(1)}_3=\kappa^{(1,1)}_3&=&(4k_2+9k_3+5k_4+2s_1-t_1)L,\nn
\beta^{(1)}_4=\kappa^{(1,1)}_4&=&3t_1L,\nn
\beta^{(1)}_5=\kappa^{(1,1)}_5&=&2\left(k_1+2k_2+k_3
+k_5+k_6-u_1\right)L,\nn
\beta^{(1)}_6=\kappa^{(1,1)}_6&=&\left(k_2+2k_3+k_4
+2k_5+5k_6+3k_7+u_1-v_1\right)L,\nn
\beta^{(1)}_7=\kappa^{(1,1)}_7&=&2v_1L,\nn
\beta^{(1)}_8=\kappa^{(1,1)}_8&=&\left(k_5+2k_6+k_7
+k_8+k_9-w_1\right)L,\nn
\beta^{(1)}_9=\kappa^{(1,1)}_9&=&w_1L.
\label{eq2}
\eea
The results adopt their simplest form for $r_1=s_1=t_1=u_1=v_1=w_1=0$, in 
which case 
the $\beta$-functions for $k_4$, $k_7$ and $k_9$ are identically
zero, and therefore the corresponding terms could be omitted from the action. 
This feature will persist at two loops; 
though it is worth pointing out that $\beta_4$, $\beta_7$ and $\beta_9$
would acquire contributions in the case of more
than one chiral field if we then included
the $yF\psi C\psi$ and $\phibar^2\psi C\psi$ interactions.
In this simple case, $r_1=s_1=t_1=u_1=v_1=w_1=0$,
we also see that once we have a 
non-zero $k_1$, we inevitably generate $k_2$ and $k_5$ and thence
$k_3$, $k_6$ and $k_8$.
We now see clearly why it was in principle necessary to give the 
$\ybar^{-1}F^3$ term its own 
coupling $k_1$ rather than $y\ybar$, since in general 
$\beta_1\ne \mu\frak{d}{d\mu}y\ybar$; and why we had to 
introduce all the other terms corresponding to $k_{2-9}$ as 
well. In fact by taking
\be
r_1=k_2, \quad s_1=t_1=u_1=v_1=w_1=0,
\ee
we can make 
$\beta^{(1)}_1$ consistent with $k_1=y\ybar$ corresponding to a coefficient
of $y$ for the $F^3$ term (as in Eq.~(\ref{defact}); but we cannot 
use the freedom
in choosing $r_1,s_1,t_1,u_1,v_1,w_1$ to maintain 
$k_{2-9}$ all zero. 

If we eliminate $F$ and $\Fbar$ from the action we find
\bea
F&=&\Gbar,\nn
\Fbar&=&G+\ybar^{-1}\left[\frak12k_1F^2+k_2F\Gbar+\frak12k_3\Gbar^2\right]
+\ybar^{-2}\mbar^2[k_5F+k_6\Gbar]+\ybar^{-3}\mbar^4k_8
\label{felim}
\eea
and the action becomes
\bea
S&=&\int d^4x\Bigl\{\pa^{\mu}\phibar 
\pa_{\mu}\phi+i\psibar\sigmabar^{\mu}\pa_{\mu}\psi\nn
&&-G\Gbar+\frak12y\phi\psi^2+\frak12\ybar\phibar\psibar^2+\frak12m\psi^2
+\frak12\mbar\psibar^2\nn
&&+\frak16\lambda_1
\ybar^{-1}\Gbar{}^3+\frak12\mbar^2\ybar^{-2}\lambda_2\Gbar{}^2
+\mbar^4\ybar^{-3}\lambda_3\Gbar\Bigr\}.
\label{selim}
\eea
where
\bea
\lambda_1&=&k_1+3(k_2+k_3)+k_4,\nn
\lambda_2&=&k_5+2k_6+k_7,\nn
\lambda_3&=&k_8+k_9.
\eea
Writing the divergent contributions to the deformed part of the effective action
 in the eliminated case as
\be
\Gamma_{C\rm{elim}}^{\rm{pole}}=-\int d^4x[Y_1\ybar^{-1}\Gbar{}^3
+\ybar^{-2}\mbar^2 Y_2\Gbar{}^2+\ybar^{-3}\mbar^4Y_3\Gbar],
\ee
(introducing an overall minus sign as in Eq.~(\ref{divdef}))
we have (using minimal subtraction)
\bea
\lambda_{1B}&=&\lambda_1+6Y_1,\nn
\lambda_{2B}&=&\lambda_2+2Y_2,\nn
\lambda_{3B}&=&\lambda_3+Y_3.
\label{lamdef}
\eea
We find from the eliminated diagrams (an example of which is
depicted in Fig.~\ref{fig1}(h)) that
\bea
Y_1^{(1)}&=&\frak52\lambda_1\frac{L}{\epsilon},\nn
Y_2^{(1)}&=&\left(\lambda_1+3\lambda_2\right)
\frac{L}{\epsilon},\nn
Y_3^{(1)}&=&\left(\lambda_2+\lambda_3\right)
\frac{L}{\epsilon},
\label{Yelim}
\eea
and then writing 
\bea
\lambda_{1B}&=&\lambda_1\sum_{n=1} \lambda_{1B}^{(n)},\nn
\lambda_{2B}&=&\mu^{-\epsilon}
\left(\lambda_2+\sum_{n=1} \lambda_{2B}^{(n)}\right)
,\nn
\lambda_{3B}&=&\mu^{-2\epsilon}\left(\lambda_3+\sum_n \lambda_{3B}^{(n)}\right),
\label{eq3}
\eea
where
\be
\lambda_{iB}^{(n)}=
\sum_{m=1}^n\frac{L^{(n,m)}_i}{\epsilon^m},
\quad i=1\ldots3,
\label{eq3a}
\ee
we find from Eqs.~(\ref{nonren}), (\ref{Gren}), (\ref{selim}),
(\ref{lamdef}), (\ref{Yelim}) that
\bea
L_1^{(1,1)}&=&15\lambda_1 L\nn
L_2^{(1,1)}&=&\left(2\lambda_1+6\lambda_2\right)L,\nn
L_3^{(1,1)}&=&\left(\lambda_2+\lambda_3 \right)L,
\label{eq4}
\eea
and then 
\be
\beta_{\lambda_i}^{(1)}=L_i^{(1,1)}.
\label{eq4a}
\ee
An important consistency check is that
\bea
\lambda_{1B}&=&k_{1B}+k_{4B}+3(k_{2B}+k_{3B}),\nn
\lambda_{2B}&=&k_{5B}+2k_{6B}+k_{7B},\nn
\lambda_{3B}&=&k_{8B}+k_{9B},
\label{consistb}
\eea
and it is easy to confirm that this is satisfied at one loop 
(irrespective of the values of $R_B,S_B,T_B,U_B,V_B,W_B$ in Eq.~(\ref{fren}))
using 
Eqs.~(\ref{eq1}), (\ref{eq1a}), (\ref{eq2}), (\ref{eq3}), (\ref{eq3a}),
(\ref{eq4}).
The original deformed Wess-Zumino action of Eq.~(\ref{defact}) 
corresponded to the values $k_1=y\ybar$, $k_{2-9}=0$. However,
as we emphasised earlier, our more general lagrangian in 
Eq.~(\ref{sunel}) is invariant under
$\Ncal=\frak12$ transformations whatever the values of $k_1-k_9$;
and we saw from Eq.~(\ref{eq2}) that the choice
$k_1=y\ybar$, $k_{2-9}=0$ is not maintained by renormalisation.
It is interesting to ask if there is any complete set of values of $k_1-k_9$ 
(or at least any form for the deformed action) 
which {\it is} preserved by renormalisation and which would be in 
some sense natural. To be precise, we ask if we can write
\bea
k_i&=&a_i (y\ybar)^{\rho},\quad i=1\ldots4,\nn
k_i&=&a_i (y\ybar)^{\sigma},\quad i=5,6,7,\nn
k_i&=&a_i (y\ybar)^{\tau},\quad i=8,9,
\eea
where $a_i$, $i=1\ldots 9$ are numbers (i.e. not functions of $y$ or $\ybar$,
and hence scale independent). This entails 
\bea
\frac{\beta_1^{(1)}}{k_1}=\frac{\beta_2^{(1)}}{k_2}&=&
\frac{\beta_3^{(1)}}{k_3}=\frac{\beta_4^{(1)}}{k_4}=
\rho\left(\frac{\beta_y^{(1)}}{y}
+\frac{\beta_{\ybar}^{(1)}}{\ybar}\right),\nn
\frac{\beta_5^{(1)}}{k_5}&=&\frac{\beta_6^{(1)}}{k_6}=
\frac{\beta_7^{(1)}}{k_7}=\sigma\left(\frac{\beta_y^{(1)}}{y}
+\frac{\beta_{\ybar}^{(1)}}{\ybar}\right),\nn
\frac{\beta_8^{(1)}}{k_8}&=&\frac{\beta_9^{(1)}}{k_9}
=\tau\left(\frac{\beta_y^{(1)}}{y}
+\frac{\beta_{\ybar}^{(1)}}{\ybar}\right).
\label{eq9}
\eea
Using Eqs.~(\ref{betaone}), (\ref{eq2}), we obtain the equations
\bea
(3-3\rho)k_1+3k_2-3r_1&=&0,\nn
4k_1+(10-3\rho)k_2+6k_3+r_1-2s_1&=&0,\nn
4k_2+(9-3\rho)k_3+5k_4+2s_1-t_1&=&0,\nn
-3\rho k_4+3t_1&=&0,\nn
2k_1+4k_2+2k_3+(2-3\sigma)k_5+2k_6-2u_1&=&0,\nn
k_2+2k_3+k_4+2k_5+(5-3\sigma)k_6+3k_7+u_1-v_1&=&0,\nn
-3\sigma k_7+2v_1&=&0,\nn
k_5+2k_6+k_7+(1-3\tau)k_8+k_9-w_1&=&0,\nn
-3\tau k_9+w_1&=&0.
\label{rgeqs}
\eea
We can solve, for instance, the first three equations in Eq.~(\ref{rgeqs})
successively for $r_1$, then $s_1$, then $t_1$. The fourth then gives the 
constraint
\be
(15-3\rho)\lambda_1=0.
\ee 
Dealing with the fifth to seventh, and eighth and ninth, equations similarly,
we also find
\bea
2\lambda_1+(6-3\sigma)\lambda_2&=&0,\nn
\lambda_2+(1-3\tau)\lambda_3&=&0.
\label{rgconds}
\eea
We then see that these conditions are equivalent to the equations we would 
have derived using Eqs.~(\ref{eq4}), (\ref{eq4a}) if we had sought a similar 
RG-invariant 
solution in the eliminated form of the theory. In other words, as was to be
expected, the eliminated form of the theory contains the same information as 
the uneliminated form.  
If, say $\lambda_1\ne0$ (which, through 
Eq.~(\ref{rgconds}) implies $\lambda_{2,3}\ne0$) then we require $\rho=5$ but 
then
we can choose $r_1$, $s_1$, $t_1$, $u_1$, $v_1$, $w_1$ to 
satisfy Eqs.~(\ref{rgeqs})
for any $k_i$; and there are similar solutions with $\lambda_1=0$, 
$\lambda_{2,3}\ne0$, $\sigma=2$ and $\lambda_1=\lambda_2=0$, $\lambda_3\ne0$,
$\tau=\frak13$, all with arbitrary $k_i$.
In the case 
\be
\lambda_1=\lambda_2=\lambda_3=0
\label{lvan}
\ee
(where the deformed potential vanishes in the eliminated case) there is 
general no constraint on $\rho$, $\sigma$, $\tau$; and for any values of 
$k_i$ satisfying
Eq.~(\ref{lvan}), and any $\rho$, $\sigma$, $\tau$, we can again choose
$r_1$, $s_1$, $t_1$, $u_1$, $v_1$, $w_1$ to satisfy Eqs.~(\ref{rgeqs}).  
However it is worth pointing out that there
are four interesting special solutions with $r_1=s_1=t_1=u_1=v_1=w_1=0$, and 
with $\rho=\sigma=\tau$, firstly
\be
k_1=\frak14k_2=\frak38k_3=\frak12k_5=\frak34k_6=3k_8, \quad k_4=k_9=0,\quad
\rho=\sigma=\tau=5,
\label{eq10}
\ee
secondly 
\bea
k_1=k_2=-\frak34k_3,\quad k_6&=&-\frak53k_1+2k_5,\quad
k_8=-\frak23k_1+k_5,\nn
k_4=k_7=k_9=0,&\quad& \rho=\sigma=\tau=2,
\label{eq9a}
\eea
thirdly
\be
k_1=-\frak32k_2=3k_3,\quad k_5=-2k_6,\quad k_4=k_7=k_9=0,\quad \rho=\sigma
=\tau=\frak13,
\label{eq8}
\ee
and fourthly
\be
k_1=-k_2=k_3=-k_4,\quad k_5=-k_6=k_7,\quad k_8=-k_9,\quad \rho=\sigma=\tau=0,
\label{eq7}
\ee
Eqs.~(\ref{eq10})-(\ref{eq7}) each correspond respectively to one of 
the general cases mentioned above.
Eqs.~(\ref{eq8}), (\ref{eq7}) are particularly intriguing since they will also 
prove to be valid in a similar way (with no non-linear renormalisation
of $\Fbar$) at two loops.
This may be related to the fact that Eqs.~(\ref{eq8}), (\ref{eq7})
correspond to actions of the simple form
\bea
S&=&\int d^4x\Bigl\{
\pa^{\mu}\phibar\pa_{\mu}\phi+i\psibar\sigmabar^{\mu}\pa_{\mu}\psi
+\Fbar F  -GF-\Gbar\Fbar\nn
&&+\frak12y\phi\psi^2+\frak12\ybar\phibar\psibar^2+\frak12m\psi^2
+\frak12\mbar\psibar^2\nn
&&+F[\frak16k_1\ybar^{-1}(F-\Gbar)^2+\frak12k_5\ybar^{-2}\mbar^2(F-\Gbar)
+k_8\ybar^{-3}\mbar^4]\Bigr\}
\label{forminva}
\eea
or
\bea
S&=&\int d^4x\Bigl\{
\pa^{\mu}\phibar\pa_{\mu}\phi+i\psibar\sigmabar^{\mu}\pa_{\mu}\psi
+\Fbar F  -GF-\Gbar\Fbar\nn
&&+\frak12y\phi\psi^2+\frak12\ybar\phibar\psibar^2+\frak12m\psi^2
+\frak12\mbar\psibar^2\nn
&&+\frak16k_1\ybar^{-1}(F-\Gbar)^3+\frak12k_5\ybar^{-2}\mbar^2(F-\Gbar)^2
+k_8\ybar^{-3}\mbar^4(F-\Gbar)\Bigr\}
\label{forminv}
\eea
respectively.  The equations of motion for $F$ and $\Fbar$ 
in Eq.~(\ref{felim}) are then 
particularly simple upon applying Eqs.~(\ref{eq8}), (\ref{eq7});
specifically, that for $F$ becomes linear (in $\Gbar$) upon   
applying $F=\Gbar$, the equation for $\Fbar$. It may also be 
significant that (again upon
applying $F=\Gbar$) these two actions are power-counting 
super-renormalisable. In this somewhat somewhat restricted sense
the values of $k_{1-9}$ in Eqs.~(\ref{eq7}), (\ref{eq8}) may be regarded
as the "natural" values we were seeking. 

It is also interesting that the values $\rho=2$ and $\rho=\frak13$ have a
significance even in the massless case, as we see in Eqs.~(\ref{eq9a}), 
(\ref{eq8}), despite these values
arising in the first instance for the coefficients 
of the massive terms in Eq.~(\ref{rgconds}).
 
\begin{figure}
\includegraphics{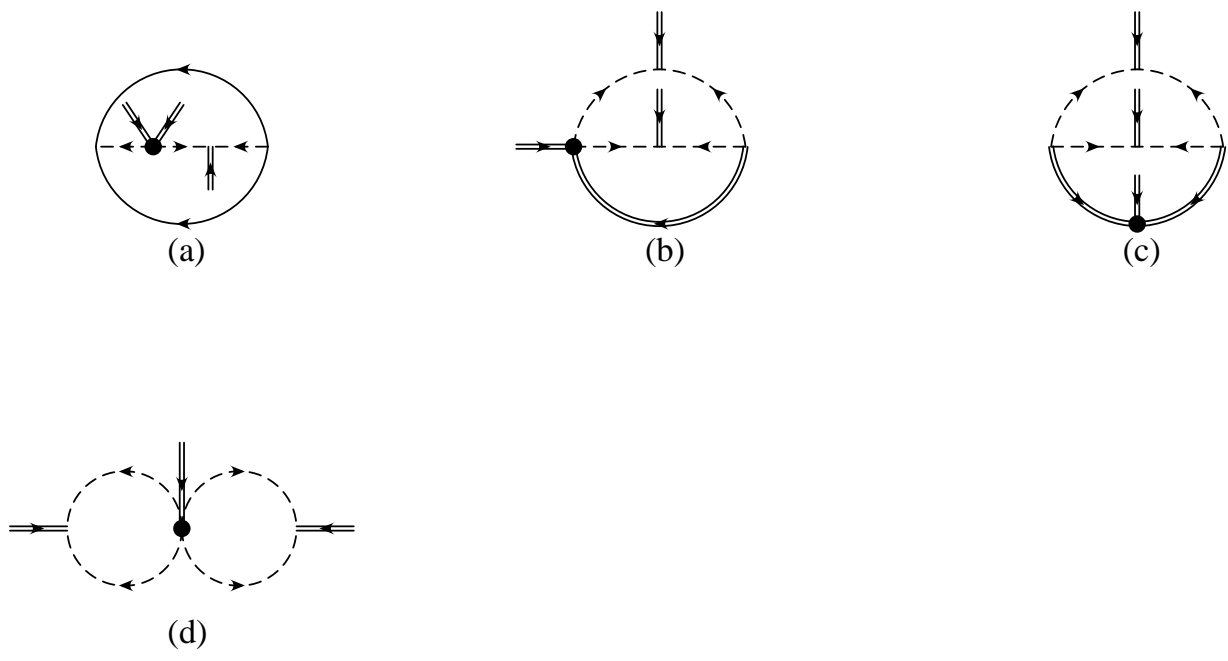}
\caption{Two-loop $F^3$ Diagrams}\label{fig2}
\end{figure}

\begin{figure}
\includegraphics{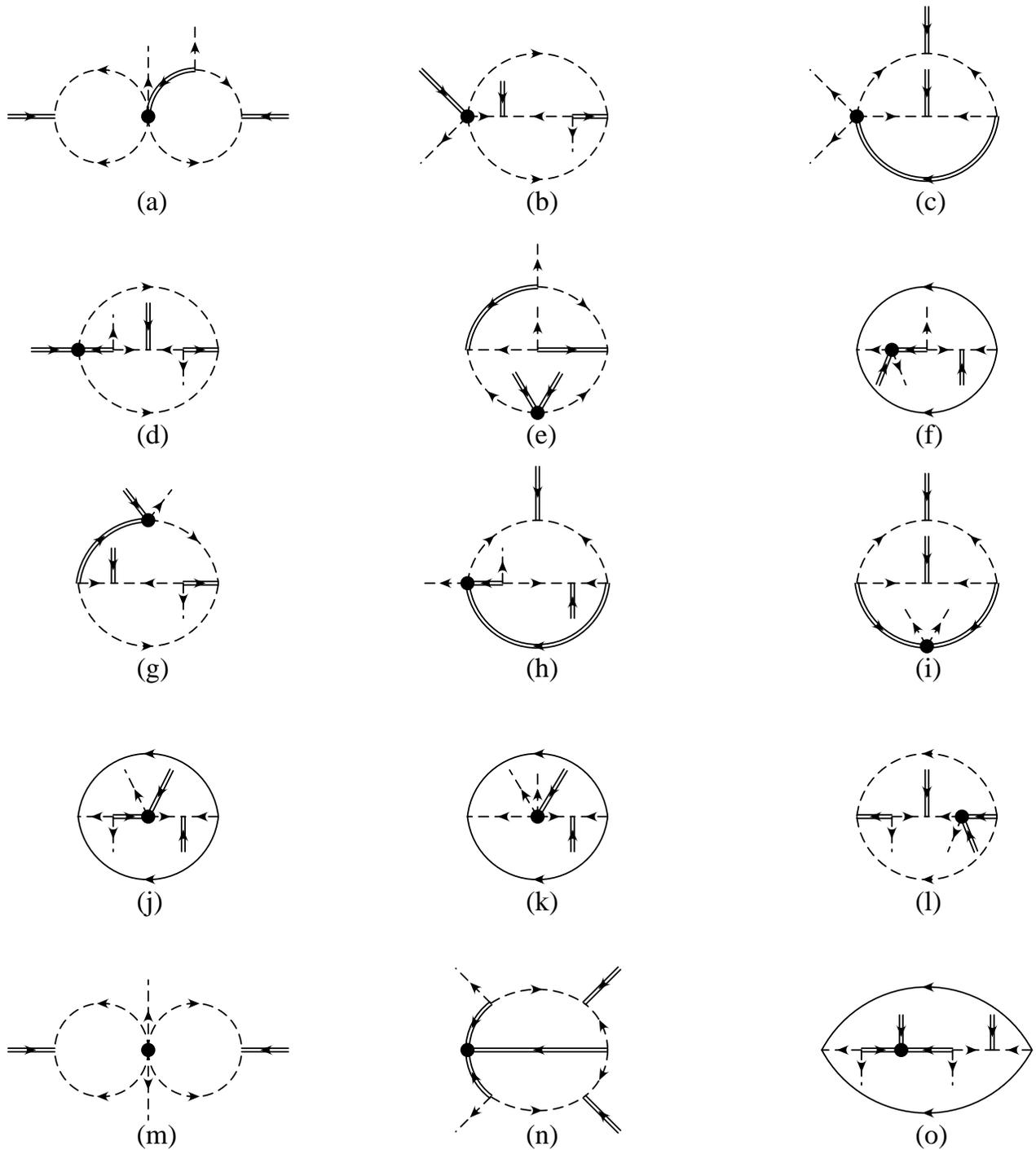}
\caption{Two-loop $F^2\phibar^2$ Diagrams} \label{fig3}
\end{figure}

\begin{figure}
\includegraphics{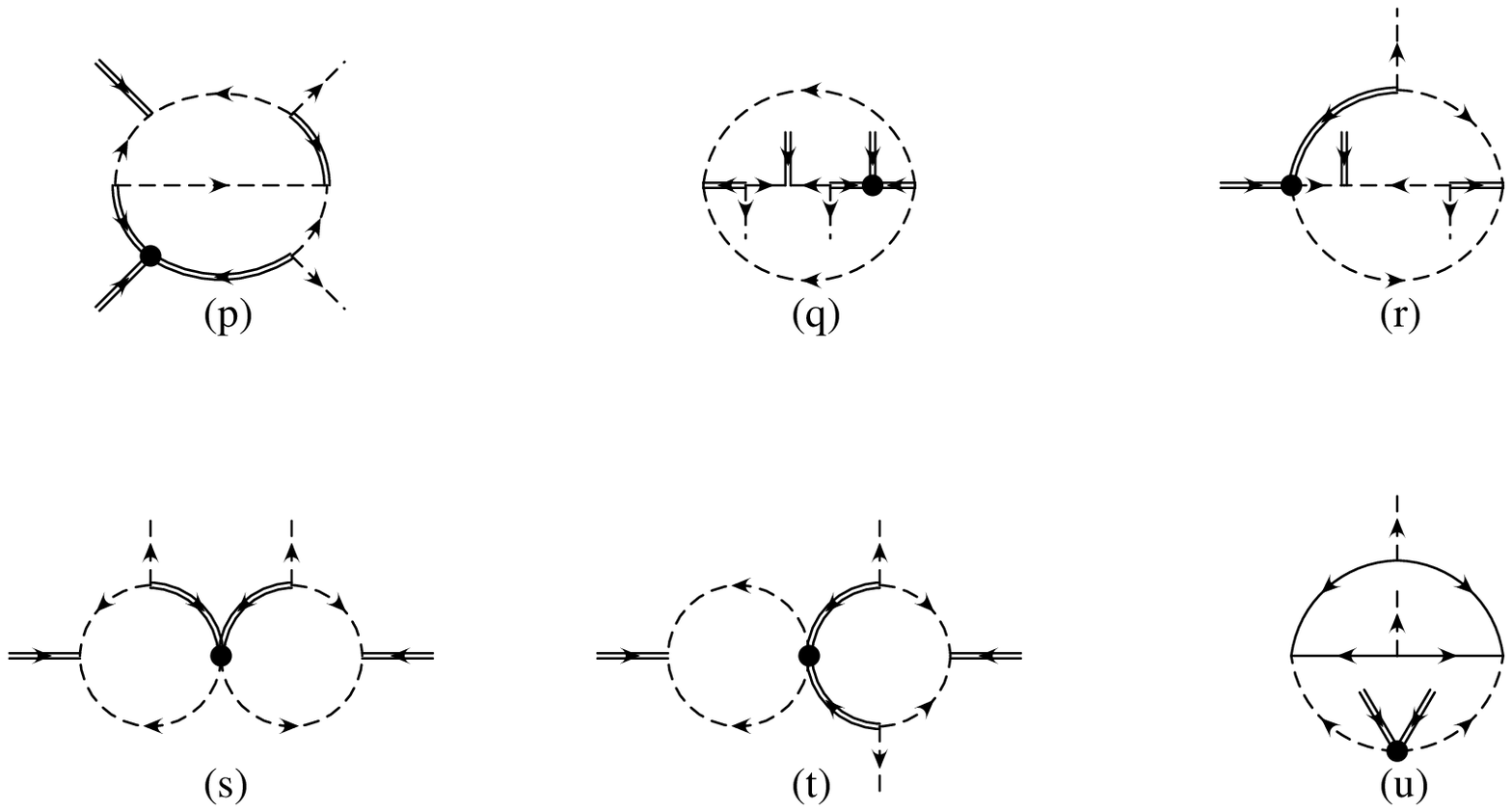}
\caption{Two-loop $F^2\phibar^2$ Diagrams (continued)}\label{fig3a}          
\end{figure}

\begin{figure}
\includegraphics{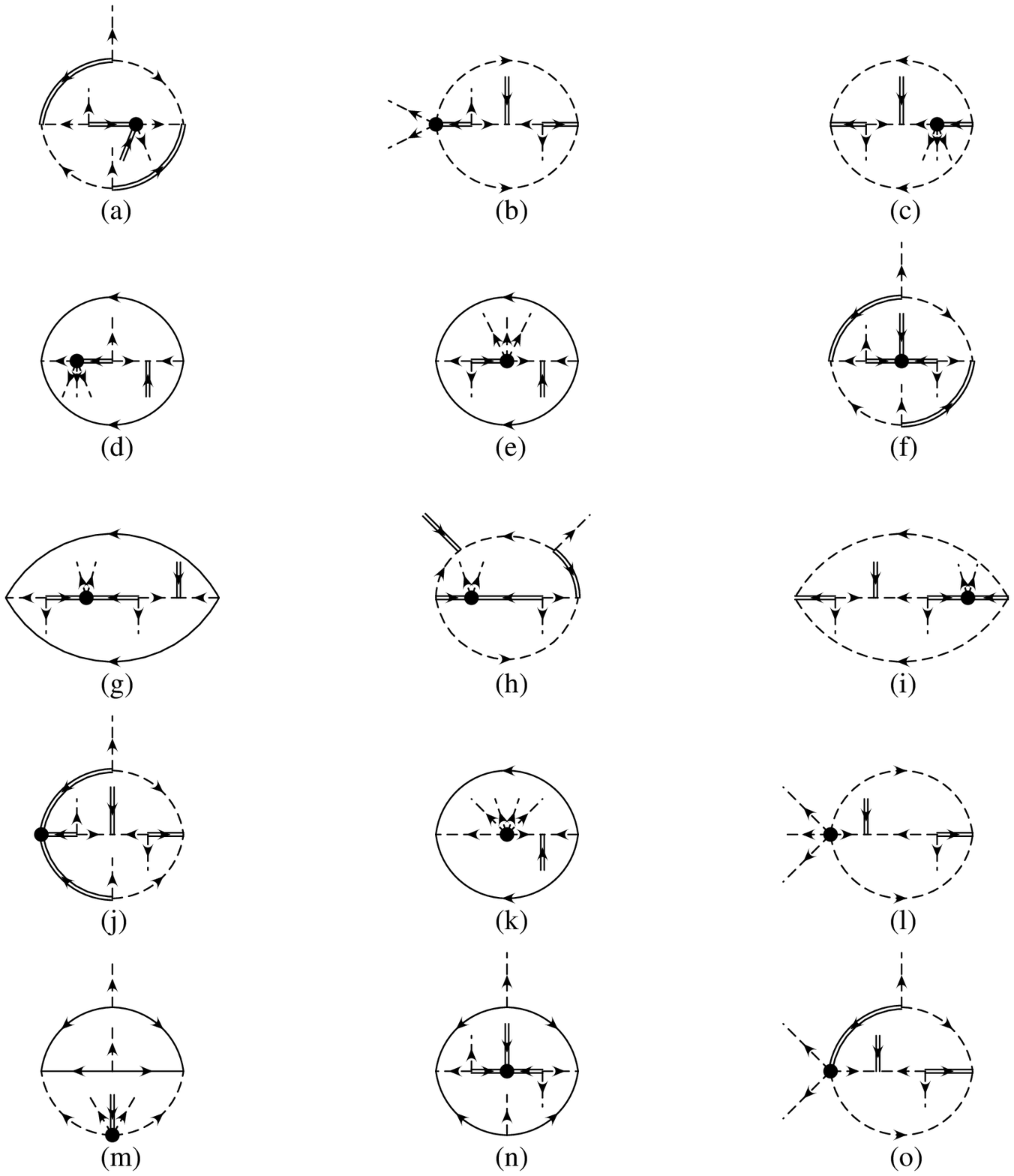}
\caption{Two-loop $F\phibar^4$ Diagrams }\label{fig4}          
\end{figure}

\begin{figure}
\includegraphics{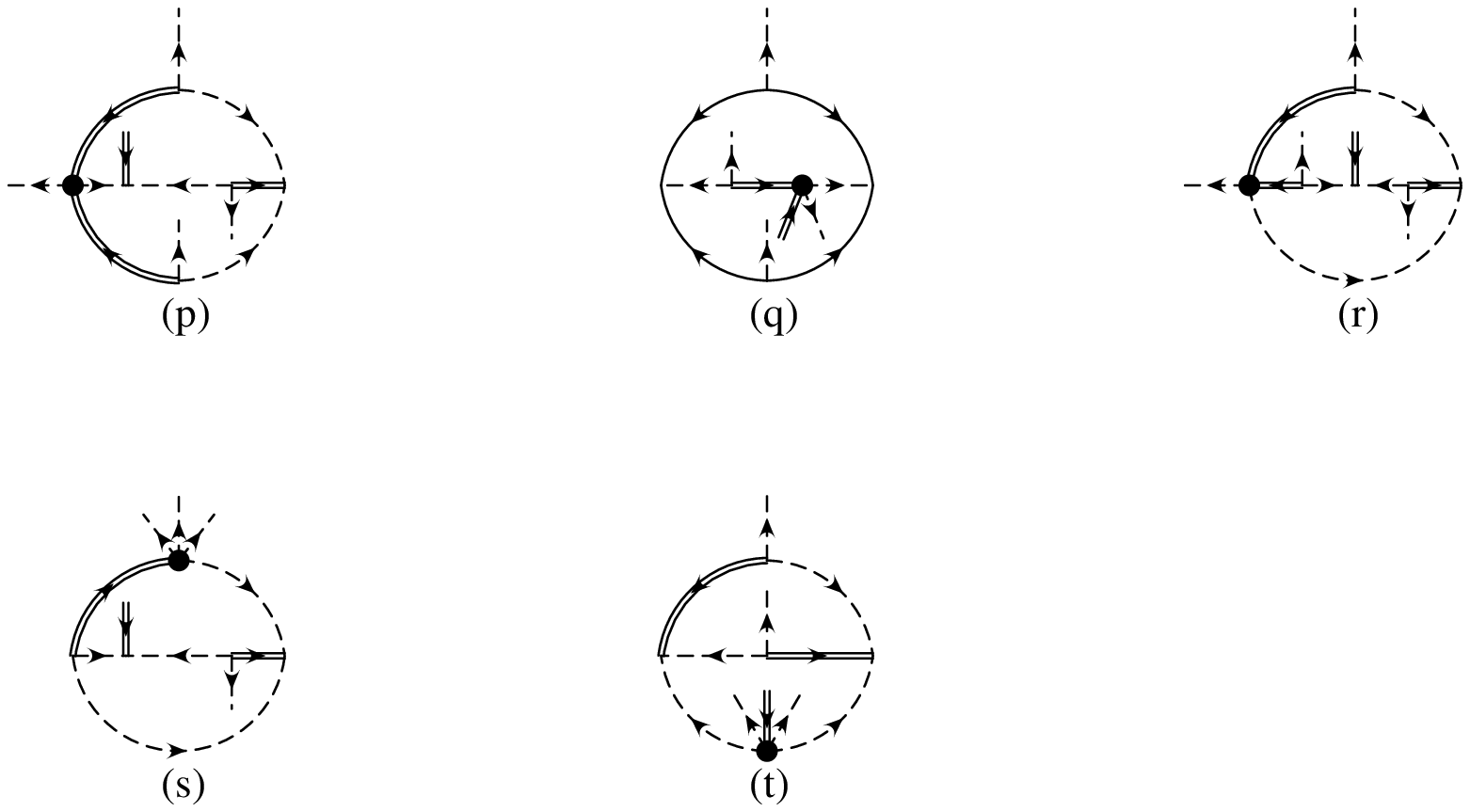}
\caption{Two-loop $F\phibar^4$ Diagrams (continued)} \label{fig4a}  
\end{figure}

\begin{figure}
\includegraphics{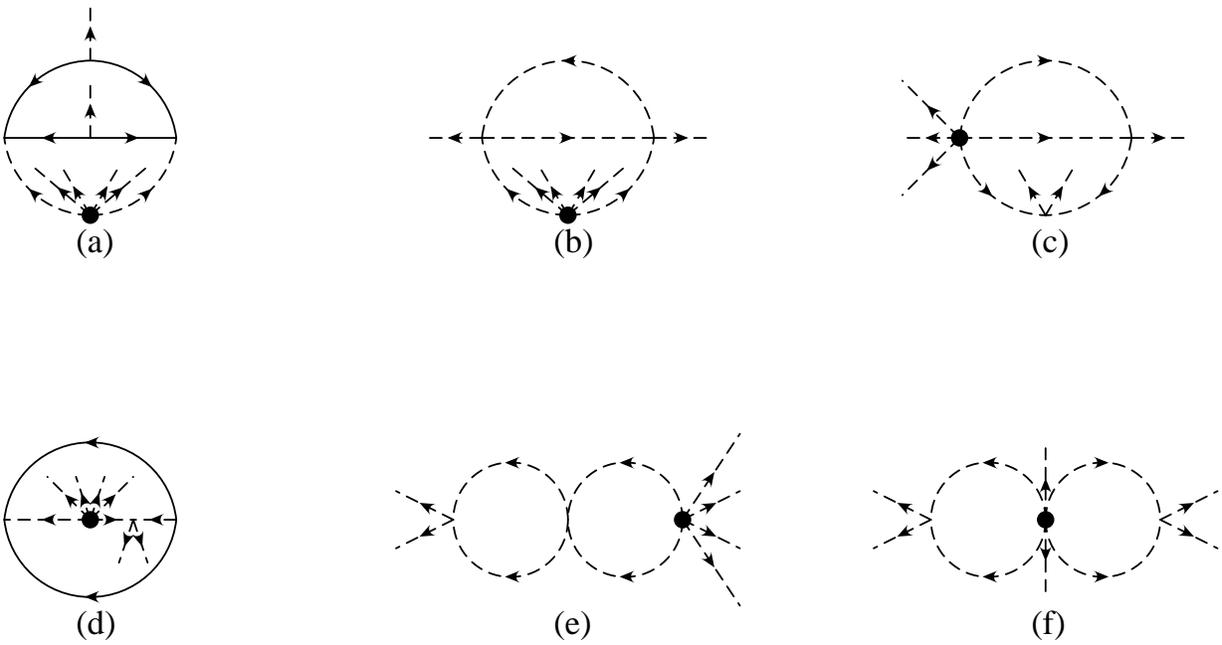}
\caption{Two-loop $\phibar^6$ Diagrams in the eliminated case}\label{fig5}  
\end{figure}

We shall now discuss the two-loop calculation which we shall see  
follows a very similar pattern. At two loops we have
\be
Z^{(2)}=-I\frac{L^2}{\epsilon^2},
\label{ztwo}
\ee
where
\be
I=1-\frac12\epsilon,
\ee
which leads (through Eq.~(\ref{nonren})) to
\be
\beta^{(2)}_y=-\frac32L^2y.
\label{betaytwo}
\ee
Examples of each group of divergent two-loop diagrams (except for those
contributing to $\Zbar_4$, $\Zbar_7$ and $\Zbar_9$) are depicted
in Figs.~\ref{fig2}-\ref{fig4a} and the divergent contributions are 
shown in Tables \ref{tabb}-\ref{tabd}
(suppressing factors of $\frak{L^2}{\epsilon^2}$). The diagrams
contributing to $\Zbar_4$, $\Zbar_7$ and $\Zbar_9$ cancel in pairs 
and in the interests of brevity are not shown explicitly; though 
in the eliminated case, a similar pair can be seen in
Figs.~\ref{fig5}(a), \ref{fig5}(b). This cancellation is due to the fact that 
for instance the diagrams contributing to $\Zbar_4$ have a one-loop 
$\phi^2\phibar{}^2$ subdiagram, and in the uneliminated case there is no 
counterterm for such a divergence.

\begin{table}
\begin{center}
\begin{tabular}{|c| c |} \hline
&$\Zbar_1$  \\ \hline
a&$\frak12Ik_2$  
\\ \hline
b&$2k_2$    $$ \\ \hline
c&$k_1$\\ \hline
d&$\frak32k_3$\\ \hline
\end{tabular}
\caption{\label{tabb} Divergent contributions from Fig.~\ref{fig2}}
\end{center}
\end{table}

\begin{table}
\begin{center}
\begin{tabular}{|c| c c |} \hline
&$\Zbar_2$&$\Zbar_5$\\ \hline
a&$12k_3$ & $6k_3$ \\ \hline
b&$6Ik_3$ & $3Ik_3$ 
\\ \hline
c&$6k_3$&$2(k_3+k_6)$ \\ \hline
d&$2Ik_2$&$Ik_2$
\\ \hline
e&$Ik_2$&$\frak12Ik_2$
\\ \hline
f&$2Ik_2$&$Ik_2$
\\ \hline
g&$4Ik_2$&$2Ik_2$
\\ \hline
h&$8k_2$&$4k_2$\\ \hline
i&$k_2$&$k_5$\\ \hline
j&$2Ik_2$&$Ik_2$
\\ \hline
k&$3Ik_3$&$I(k_3+k_6)$  \\ \hline
l&$2Ik_2$&$Ik_2$
\\ \hline
m&$\frak{15}{2}k_4$&$\frak32(2k_4+k_7)$\\ \hline
n&$4k_1$&$2k_1$\\ \hline
o&$2Ik_1$&$Ik_1$
\\ \hline
p&$4Ik_1$&$2Ik_1$\\ \hline
q&$2Ik_1$&$Ik_1$
\\ \hline
r&$4Ik_2$&$2Ik_2$
\\ \hline
s&$4k_2$&$2k_2$\\ \hline
t&$2k_2$&$k_2$\\ \hline
u&$-Ik_2$&$-\frak12Ik_2$
\\ \hline
\end{tabular}
\caption{\label{tabc} Divergent contributions from Fig.~\ref{fig3}}
\end{center}
\end{table}

\begin{table}
\begin{center}
\begin{tabular}{|c| c c c |} \hline
&$\Zbar_3$&$\Zbar_6$&$\Zbar_8$  \\ \hline
a&$8Ik_2$ & $8Ik_2$  & $2Ik_2
$ \\ \hline
b&$6Ik_3$ & $I(5k_3+2k_6)$  & $I(k_3+k_6)$$$ \\ \hline
c&$2Ik_3$&$I(k_3+2k_6)$&$Ik_6 $\\ \hline
d&$2Ik_3$&$I(k_3+2k_6)$&$Ik_6 $\\ \hline
e&$2Ik_3$&$I(k_3+2k_6)$&$Ik_6 $\\ \hline
f&$4Ik_1$&$4Ik_1$&$Ik_1
$\\ \hline
g&$2Ik_2$&$I(k_2+2k_5)$&$Ik_5 $\\ \hline
h&$4Ik_2$&$2I(k_2+2k_5)$&$2Ik_5 $\\ \hline
i&$2Ik_2$&$I(k_2+2k_5)$&$Ik_5 $\\ \hline
j&$4Ik_1$&$4Ik_1$&$Ik_1$
\\ \hline
k&$\frak52Ik_4$&$I(k_4+3k_7)$&$I(k_9+k_7) $ \\ \hline
l&$10Ik_4$&$I(7k_4+6k_7)$&$I(k_4+3k_7)$\\ \hline
m&$-6Ik_3$&$-I(5k_3+2k_6)$&$-I(k_3+k_6)$\\ \hline
n&$-4Ik_1$&$-4Ik_1$&$-Ik_1$
\\ \hline
o&$12Ik_3$&$2I(5k_3+2k_6)$&$2I(k_3+k_6)$\\ \hline
p&$4Ik_2$&$4Ik_2$&$Ik_2$\\ \hline
q&$-8Ik_2$&$-8Ik_2$
&$-2Ik_2$\\ \hline
r&$8Ik_2$&$8Ik_2$&$2Ik_2$
\\ \hline
s&$4Ik_3$&$2I(k_3+2k_6)$&$2Ik_6 $\\ \hline
t&$6Ik_3$&$I(5k_3+2k_6)$&$I(k_3+k_6)$\\ \hline
\end{tabular}
\caption{\label{tabd} Divergent contributions from Fig.~\ref{fig4}}
\end{center}
\end{table}

\begin{table}
\begin{center}
\begin{tabular}{|c| c c c|} \hline
&$\Gbar{}^3$&$\mbar^2\Gbar{}^2$&$\mbar^4\Gbar$  \\ \hline
a&$-5I\lambda_1$ & $-\frak{9}{2}I\lambda_1-6I\lambda_2$  
& $-I(\lambda_1+5\lambda_2+2\lambda_3)$\\ \hline
b&$5I\lambda_1$ & $\frak{9}{2}I\lambda_1+6I\lambda_2$  
& $I(\lambda_1+5\lambda_2+2\lambda_3)$ \\ \hline
c&$10I\lambda_1$&$7I\lambda_1+6I\lambda_2$&$I(\lambda_1+3\lambda_2)$ \\ \hline
d&$\frak52I\lambda_1$&$I\lambda_1+3I\lambda_2$&$I(
\lambda_2+\lambda_3)$\\ \hline
e&$\frak52\lambda_1$&$\lambda_1+3\lambda_2$&$(
\lambda_2+\lambda_3)$ \\ \hline
f&$\frak{15}{2}\lambda_1$&$3\lambda_1+\frak32\lambda_2$&
$0$\\ \hline
\end{tabular}
\caption{\label{tabe} Divergent contributions from Fig.~\ref{fig5}}
\end{center}
\end{table}

The total two-loop divergences
are given by
\bea
\Zbar_1^{(2)}&=&\frac{L^2}{\epsilon^2}\left\{k_1+[\frak12I+2]k_2
+\frak32k_3\right\},\nn
\Zbar_2^{(2)}&=&\frac{L^2}{\epsilon^2}\left\{
4[2I+1]k_1+[16I+15]k_2+
9[I+2]k_3+\frak{15}{2}k_4\right\},\nn
\Zbar_3^{(2)}&=&\frac{L^2}{\epsilon^2}\left\{
[4k_1+20k_2+28k_3+\frak{25}{2}k_4]I\right\},\nn
\Zbar_4^{(2)}&=&0,\nn
\Zbar_5^{(2)}&=&\frac{L^2}{\epsilon^2}\Biggl\{
2\left([2I+1]k_1
+\left[4I+\frak72\right]k_2
+2[I+2]k_3+\frak32k_4\right)\nn
&&+k_5+[I+2]k_6+\frak32k_7\Biggr\},\nn
\Zbar_6^{(2)}&=&\frac{L^2}{\epsilon^2}\Bigl\{
2\left(2k_1+8k_2+10k_3+4k_4\right)
+8k_5+16k_6+9k_7\Bigr\}I,\nn
\Zbar_7^{(2)}&=&0,\nn
\Zbar_8^{(2)}&=&\frac{L^2}{\epsilon^2}\Bigl\{
k_1+3k_2+3k_3+k_4
+4\left(k_5+2k_6+k_7\right)
+k_9\Bigr\}I,\nn
\Zbar_9^{(2)}&=&0.
\label{ztwoa}
\eea
The two-loop contributions to the bare couplings are given (from the bare form
of Eq.~(\ref{sunel}) and using Eqs.~(\ref{nonren}), (\ref{Gren})) by
\bea
k_{1B}^{(2)}&=&6\Zbar_1^{(2)}-3Z^{(2)}k_1-3\left(Z^{(1)}\right)^2k_1
-3Z^{(1)}\left(k^{(1)}_{1B}+3R^{(1)}_{B}\right)-3R^{(2)}_{B},\nn
k_{2B}^{(2)}&=&2\Zbar_2^{(2)}-2Z^{(2)}k_2
-\left(Z^{(1)}\right)^2k_2-2Z^{(1)}\left(k^{(1)}_{2B}-R^{(1)}_{B}
+2S^{(1)}_{B}\right)+R^{(2)}_{B}-2S^{(2)}_{B},\nn
k_{3B}^{(2)}&=&2\Zbar_3^{(2)}-Z^{(2)}k_3
-Z^{(1)}\left(k^{(1)}_{3B}-2S^{(1)}_{B}+T^{(1)}_{B}\right)+2S^{(2)}_{B}
-T^{(2)}_{B},\nn
k_{4B}^{(2)}&=&6\Zbar_4^{(2)}+3T^{(2)}_{B},\nn
k_{5B}^{(2)}&=&2\Zbar_5^{(2)}-2Z^{(2)}k_5-\left(Z^{(1)}\right)^2k_5
-2Z^{(1)}\left(k_{5B}^{(1)}+2U^{(1)}_{B}\right)-2U^{(2)}_{B},\nn
k_{6B}^{(2)}&=&\Zbar_6^{(2)}-Z^{(2)}k_6
-Z^{(1)}\left(k_{6B}^{(1)}-U^{(1)}_{1B}+V^{(1)}_{B}\right)
+U^{(2)}_{B}-V^{(2)}_{B},\nn
k_{7B}^{(2)}&=&2\Zbar_7^{(2)}+2V^{(2)}_{B},\nn
k_{8B}^{(2)}&=&\Zbar_8^{(2)}-Z^{(2)}k_8-Z^{(1)}\left(k_{8B}^{(1)}
+W^{(1)}_{B}\right)-W^{(2)}_{B},\nn
k_{9B}^{(2)}&=&\Zbar_9^{(2)}+W^{(2)}_{B},
\label{ktwoa}
\eea
which yields, using Eqs.~(\ref{eq1}), (\ref{eq2}), (\ref{ztwo}), (\ref{ztwoa}),
\bea
k_{1B}^{(2)}&=&[3(I+4)k_1+3(I+7)k_2+9k_3]\frac{L^2}{\epsilon^2}-3R^{(2)}_B,\nn
k_{2B}^{(2)}&=&[16(I+1)k_1+(34I+49)k_2+6(3I+8)k_3+15k_4]\frac{L^2}{\epsilon^2}
+R^{(2)}_B-2S^{(2)}_B,\nn
k_{3B}^{(2)}&=&[8Ik_1+4(10I+1)k_2+3(19I+3)k_3+5(5I+1)k_4]
\frac{L^2}{\epsilon^2}+2S^{(2)}_B-T^{(2)}_B,\nn
k_{4B}^{(2)}&=&3T^{(2)}_B,\nn
k_{5B}^{(2)}&=&[8(I+1)k_1+2(8I+11)k_2+4(2I+5)k_3+6k_4+(2I+5)k_5+2(I+4)k_6
+3k_7]\frac{L^2}{\epsilon^2}\nn
&&-2U^{(2)}_B,\nn
k_{6B}^{(2)}&=&[4Ik_1+(16I+1)k_2+2(10I+1)k_3+(8I+1)k_4+2(4I+1)k_5\nn
&&+(17I+5)k_6
+3(3I+1)k_7]\frac{L^2}{\epsilon^2}+U^{(2)}_B-V^{(2)}_B,\nn
k_{7B}^{(2)}&=&2V^{(2)}_B,\nn
k_{8B}^{(2)}&=&[I(k_1+3k_2+3k_3+k_4)+(4I+1)(k_5+2k_6+k_7)+(I+1)(k_8+k_9)]
\frac{L^2}{\epsilon^2}-W^{(2)}_B,\nn
k_{9B}^{(2)}&=&W^{(2)}_B.
\label{ktwo}
\eea
As at one loop, requiring that $k_{iB}$ as given by 
Eqs.~(\ref{eq1}), (\ref{eq1a}) be $\mu$-independent gives 
\be
\beta_i^{(2)}=2\kappa^{(2,1)}
\label{betatwo}
\ee
together with the consistency conditions for 
the two-loop double poles, 
\bea
2(16\pi^2)\kappa_{1}^{(2,2)}&=&3(\kappa^{(1,1)}_1+\kappa^{(1,1)}_2)y\ybar
+3\left(\beta_y^{(1)}\ybar
+y\beta_{\ybar}^{(1)}\right)(k_1+k_2),\nn
2(16\pi^2)\kappa_{2}^{(2,2)}&=&2(2\kappa^{(1,1)}_1+5\kappa^{(1,1)}_2
+3\kappa^{(1,1)}_3)y\ybar
\nn
&&+2\left(\beta_y^{(1)}\ybar
+y\beta_{\ybar}^{(1)}\right)(2k_1+5k_2+3k_3),\quad \hbox{etc.}
\label{constwo}
\eea
From Eqs.~(\ref{eq1}), (\ref{eq1a}), (\ref{ktwo}), (\ref{betatwo}),  the 
two-loop $\beta$-functions are hence given by
\bea
\beta_1^{(2)}&=&-3\left(k_1+k_2+2r_2\right)L^2,\nn
\beta_2^{(2)}&=&-2\left(8k_1+17k_2+9k_3-r_2+2s_2\right)L^2,\nn
\beta_3^{(2)}&=&-\left(8k_1+40k_2+57k_3+25k_4-4s_2+2t_2\right)L^2,\nn
\beta_4^{(2)}&=&6t_2L^2,\nn
\beta_5^{(2)}&=&-2\left(4[k_1+2k_2+k_3]+k_5+k_6+2u_2\right)L^2,\nn
\beta_6^{(2)}&=&-\left(4k_1+16k_2+20k_3+8k_4
+8k_5+17k_6+9k_7-2u_2+2v_2\right)L^2,\nn
\beta_7^{(2)}&=&4v_2L^2,\nn
\beta_8^{(2)}&=&-(k_1+3k_2+3k_3+k_4
+4[k_5+2k_6+k_7]+k_8+k_9+2w_2)L^2,\nn
\beta_9^{(2)}&=&2w_2L^2,
\eea
(writing $r^{(2,1)}=r_2L^2$, etc),
and we can check using Eqs.~(\ref{eq1}), (\ref{eq1a}), (\ref{betaone}),  
(\ref{eq2}), (\ref{ktwo}), that Eq.~(\ref{constwo}) is satisfied, provided we
take
\bea
r^{(2,2)}&=&(r_1+s_1)L^2,\nn
s^{(2,2)}&=&\frak12(r_1+5s_1+3t_1)L^2,\nn
t^{(2,2)}&=&0,\nn
u^{(2,2)}&=&\frak12(r_1+2s_1+t_1+u_1+v_1)L^2,\nn
v^{(2,2)}&=&0,\nn
w^{(2,2)}&=&0.
\eea
Of course due to the arbitrariness of the non-linear renormalisations 
of $\Fbar$ there is no obvious way of verifying these double pole relations
by direct calculation.

Examples of the classes of two-loop diagrams in the eliminated case are 
depicted in Fig.~\ref{fig5}, 
and the corresponding divergent contributions are listed in Table \ref{tabe}.
We find
\bea
Y_1^{(2)}&=&\frac{L^2}{\epsilon^2}\left[\frak{25}{2}I
+10\right]\lambda_1,\nn
Y_2^{(2)}&=&\frac{L^2}{\epsilon^2}\left[2I
+1\right]\left[4\lambda_1+\frak92\lambda_2\right],\nn
Y_3^{(2)}&=&\frac{L^2}{\epsilon^2}\left\{
I\lambda_1+\left[4I+1\right]
\lambda_2+\left[I+1\right]\lambda_3\right\},
\label{Ytwo}
\eea
As usual, requiring $\mu$-independence of $\lambda_{iB}$ in Eq.~(\ref{eq3})
leads to
\be
\beta_{\lambda_i}^{(2)}=2L_i^{(2,1)}
\ee
and the consistency conditions for two-loop double poles
\bea
2(16\pi^2)L_1^{(2,2)}&=&15\left[L_1^{(1,1)}y\ybar+\left(\beta_y^{(1)}\ybar
+y\beta_{\ybar}^{(1)}\right)\lambda_1\right],\nn
2(16\pi^2)L_2^{(2,2)}&=&\left(2L_1^{(1,1)}+6L_2^{(1,1)}\right)y\ybar
+\left(\beta_y^{(1)}\ybar
+y\beta_{\ybar}^{(1)}\right)(2\lambda_1+6\lambda_2),\nn
2(16\pi^2)L_3^{(2,2)}&=&\left(L_2^{(1,1)}+L_3^{(1,1)}\right)y\ybar
+\left(\beta_y^{(1)}\ybar     
+y\beta_{\ybar}^{(1)}\right)(\lambda_2+\lambda_3).
\label{consista}
\eea
This leads (via Eqs.~(\ref{lamdef}),
(\ref{eq3}), (\ref{eq3a}), (\ref{Ytwo})) to $\beta$-functions
\bea
\beta^{(2)}_{\lambda_1}&=&-75\lambda_1L^2,\nn
\beta^{(2)}_{\lambda_2}&=&-2(8\lambda_1+9\lambda_2)L^2,\nn
\beta^{(2)}_{\lambda_3}&=&-(\lambda_1+4\lambda_2+\lambda_3)L^2,
\label{elimbettwo}
\eea
and we can check using Eqs.~(\ref{betaone}), (\ref{lamdef}),
(\ref{eq3}), (\ref{eq3a}), (\ref{eq4}), (\ref{Ytwo}), 
that Eq.~(\ref{consista}) is satisfied.

As at one loop, Eq.~(\ref{consistb}) is crucial for consistency between the
uneliminated and eliminated formalisms, and leads to 
(using Eqs.~(\ref{kone}), (\ref{lamdef}), (\ref{ktwoa}))
\bea
\Zbar^{(2)}_1+\Zbar^{(2)}_2+\Zbar^{(2)}_3+\Zbar^{(2)}_4-\frak12Z^{(2)}(k_1+2k_2+k_3)&\nn
-Z^{(1)}\left(3\Zbar_1^{(1)}+2\Zbar_2^{(1)}+\Zbar_3^{(1)}\right)
+\frak12\left(Z^{(1)}\right)^2(2k_1+3k_2+k_3)&=&\frak16\lambda^{(2)}_{1B},\nn
\Zbar^{(2)}_5+\Zbar^{(2)}_6+\Zbar^{(2)}_7-Z^{(2)}(k_5+k_6)
-Z^{(1)}(2\Zbar_5^{(1)}+\Zbar_6^{(1)})&\nn+\frak12\left(Z^{(1)}\right)^2(3k_5+2k_6)
&=&\frak12\lambda^{(2)}_{2B},\nn
\Zbar_8^{(2)}+\Zbar_9^{(2)}-Z^{(2)}k_8-Z^{(1)}\Zbar_8^{(1)}
+\left(Z^{(1)}\right)^2k_8&=&\lambda_{3B}^{(2)}.
\label{consist}
\eea
It is easy to check with the aid of Eqs.~(\ref{zone}), (\ref{zonea}), 
(\ref{ztwo}), (\ref{ztwoa}), (\ref{Ytwo}), that this is satisfied.

As at one loop, if we solve Eq.~(\ref{eq9}) we find consistency conditions,
in this case 
\bea
(75+3\rho)\lambda_1&=&0\nn
16\lambda_1+(18+3\sigma)\lambda_2&=&0,\nn
\lambda_1+4\lambda_2+(1+3\tau)\lambda_3&=&0.
\eea
Once again we 
see that these conditions are equivalent to the equations we would 
have derived using Eqs.~(\ref{elimbettwo}) if we had sought a similar 
RG-invariant 
solution in the eliminated form of the theory. 
Moreover,
although as emphasised previously 
these are the only constraints on $k_{1-9}$, $\rho$, 
$\sigma$ and $\tau$, we see using Eq.~(\ref{betaytwo})
that if we impose either Eq.~(\ref{eq7}) or Eq.~(\ref{eq8})
then Eq.~(\ref{eq9}) is once again satisfied at two loops 
with $r_2=s_2=t_2=u_2=v_2=w_2=0$. It it intriguing that the values 
in Eqs.~(\ref{eq7}), (\ref{eq8}) 
are singled out at both one and two loops. 

Finally a few words about the differences between our calculation and that of 
Ref.~\cite{grisa}. The authors of Ref.~\cite{grisa} identify the coefficient 
of $(\det C) F^3$ with the Yukawa coupling (which they denote as $g$),
since that is the result of casting the deformed 
classical superspace action into its component form.
They also introduce $k'_1\mbar^4F$ and 
$k'_2\mbar^2F^2$ terms (we have added the prime to distinguish from our
own $k_1$, $k_2$ which have a different meaning). 
Therefore, we should be able to read off their results for individual diagrams 
from Tables \ref{taba}-\ref{tabd} by writing 
\bea
k_1\rightarrow 2g\gbar(\det C) ,&\quad& k_5\rightarrow 8\gbar^2k'_2(\det C) , 
\quad
k_8\rightarrow 8\gbar^3k'_1(\det C), \nn
\epsilon\rightarrow2\epsilon, &\quad&
y\rightarrow 2g,\quad \ybar\rightarrow2\gbar
\label{subst}
\eea
(including appropriate adjustments for our differing conventions), setting
the remaining $k_i$ to zero, and remembering the factors of
$\ybar$ in Eq.~(\ref{divdef}). Indeed, our results then agree precisely with 
those from their Eq.~(5.10). However, the difference between our 
results and theirs appears in the two-loop $\beta$-functions. 
Both we and they 
eventually derive results in which $F$ and $\Gbar$ are effectively identified.
Taking account of the differences in our definitions, we might expect 
that our eliminated results would be equivalent to theirs under  
\be
\lambda_1\rightarrow g\gbar\gamma,
\label{comp}
\ee
where their $\gamma$ is a dimensionless coupling associated with $|C|^2F^3$.
However, it is clear that the $\beta$-functions for our $\lambda_1$ and their 
$\gamma$ are in agreement at one loop but not at two loops. Of course 
it is well-known that $\beta$-functions are scheme-dependent (i.e.
dependent on the renormalisation used) beyond one loop in general. The change 
from one scheme to another may be effected by a redefinition of the couplings
of the theory. Nevertheless, we do not believe that the difference betwen our 
two-loop $\beta$-function and that of Ref.~\cite{grisa} 
is a consequence of using different renormalisation schemes;
both we and they are using dimensional regularisation and indeed 
as we mentioned, our results for individual diagrams agree precisely with 
theirs, confirming that they correspond to the same renormalisation scheme.

In fact, the difference resides in the precise way of identifying $F$
and $\Gbar$.
In our case this is through eliminating $F$ and $\Fbar$ via their
equations of motion Eq.~(\ref{felim}).
One might worry that this equation is only valid at the classical level; but 
we have shown in detail that we get the same results by eliminating $F$ 
and $\Fbar$ at
the classical level (hence using the classical version of Eq.~(\ref{felim}))
and then renormalising the eliminated theory; or
renormalising the uneliminated theory and then eliminating $F$ and $\Fbar$
in the 
renormalised theory (hence using the bare version of 
Eq.~(\ref{felim})). 
This equivalence is expressed by Eq.~(\ref{consistb}).
There seems little doubt therefore that our procedure is
consistent.
Moreover there is ample evidence in the literature 
(at least in the undeformed case)
that the same results (for renormalisation constants, $\beta$-functions etc)
hold in the eliminated theory obtained by 
applying the classical equations of motion, as for the uneliminated theory
(where calculations are usually performed in superfields of course).
The renormalisation Eq.~(\ref{fren}) could be regarded
as leading to a quantum modification of the equations of motion for $F$, but
we have seen that it has no effect upon the eliminated theory. 

On the other hand, the authors of Ref.~\cite{grisa}
identify $\Gbar$ and $F$ through a different, and inequivalent, process, which 
involves reassessing the identification diagrammatically at each loop level.  
It seems possible that both approaches are  
internally consistent, but that the eliminated theories thereby obtained are 
simply different at the quantum level. 
Both approaches lead to consistent two-loop double poles, at least as far
as has been computed; it is only the simple poles (and hence the 
$\beta$-functions) which differ.
Let us explain in detail how this happens (and from now on
for ease of comparison we shall describe the procedure of Ref.~\cite{grisa}
using our own notation). 
Since these authors  
identify $F$ and $\Gbar$ immediately, 
they do not introduce separate couplings
for (for instance) $F\Gbar{}^2$, $F^2\Gbar$, $\Gbar{}^3$ in order to
cancel all the different divergent terms; instead they identify
all these terms with $F^3$ (up to factors) and omit $k_{2-4}$. The  
subtlety is that they identify $\Gbar{}^2$ with 
$\frak12 F^2$ (and presumably $\Gbar{}^3$ with $\frak16 F^3$). This means that 
they effectively make the replacement
\bea
\Zbar_1&\rightarrow& \Zbar_1+\Zbar_2+\frak12\Zbar_3+\frak16\Zbar_4,\nn
\Zbar_i&\rightarrow&0, \quad i=2,3,4,\nn
k_i&\rightarrow&0, \quad i=2,3,4.
\label{substa}
\eea
Now in the ``eliminated'' case, both we and they agree that the double poles
in $\lambda_{1B}^{(2)}$ are as given by Eq.~(\ref{consista}) (equivalent to 
their Eq.~(6.5)). On the other hand, since they are starting from the  
uneliminated calculation, they are deriving $\lambda_{1B}^{(2)}$ by assuming 
Eq.~(\ref{consistb}), and therefore 
their double poles in $\lambda_{1B}^{(2)}$ are given (as are ours) according 
to Eq.~(\ref{consist}) (but in their case, after applying
Eq.~(\ref{substa})). How can we and they obtain the same double poles despite 
using different $\Zbar_i$ in Eq.~(\ref{consist})?
It happens as follows: effectively, through Eq.~(\ref{substa}) they have 
reduced $Z^{(2)}_3$ in Eq.~(\ref{consist}) by $\frak12$ ($Z^{(2)}_4$ 
is zero in this comparison since we are setting $k_{2-4}=0$); and meanwhile 
the $2k_1$ from $\Zbar_2^{(1)}$ in Eq.~(\ref{zonea}) ($\Zbar_3^{(1)}$,
$\Zbar_4^{(1)}$ being zero
for $k_{2-4}=0$) is effectively 
transferred to $\Zbar_1^{(1)}$ which, crucially, has a different factor 
in Eq.~(\ref{consist}) from $\Zbar_2^{(1)}$. 
It turns out that these two alterations have no 
effect on the double pole in $\lambda_{1B}^{(2)}$, but change the simple pole 
and hence the two-loop $\beta$-function for $\lambda_1$ (or equivalently 
$\gamma$). However it should be said that their different means of identifying
$F$ and $\Gbar$ at higher loops means that the identification of
$\lambda_1$ and $\gamma$ is perhaps problematic at higher loops;
and there is therefore 
no clear translation between our results and theirs beyond one loop.

\section{Conclusions}
We have performed a complete analysis up to two loops of the 
renormalisation of the non-anticommutative Wess-Zumino model.
We have shown that in the uneliminated case it is necessary to include
all the possible terms which can be generated by renormalisation with their
own couplings, and that this leads to results equivalent to those obtained
in the eliminated theory. In particular, if one seeks renormalisation-group
invariant trajectories for the couplings in the uneliminated theory, one obtains
essentially the same solutions as in the eliminated theory; although there
are two interesting special solutions for which the uneliminated action
adopts a simple form and which require no non-linear 
renormalisation of the auxiliary fields, at least up to two loops.
It would be interesting to see if this behaviour persists to all orders; and 
also to perform a similar analysis for a gauged model. 

In the wider context,
some of the earliest investigations of supersymmetry were motivated by
the hope that  theories might be found which were non-renormalisable by
naive power counting and yet nevertheless renormalisable  in requiring
only a finite number of local counter-terms to create a UV-finite
effective action.  Thanks to the pseudo-symmetries Eqs.~(\ref{psea}),
(\ref{pseb}) the theory
studied here (and indeed its generalisation to a gauge theory) provides
an explicit realisation of this phenomenon. It is tempting (but
presumably fanciful) to speculate  on a connection with the recent
suggestions\cite{bern}
 that ${\cal N} =8$ supergravity  (similarly naively
non-renormalisable) might in fact be finite. In any event we believe these 
theories deserve further investigation.

\vspace*{1em}

\noindent
{\large\bf Acknowledgements}\\
RP was supported by STFC through a graduate studentship. DRTJ was visiting
the Aspen Center for Physics and CERN
while part of this work was carried out, and gratefully acknowledges financial
support from CERN.

\end{document}